\documentclass[useAMS,usenatbib]{mn2e}
\usepackage{graphicx}
\usepackage{color}
\usepackage{amsmath}
\usepackage{lscape} 
\usepackage{times} 

\def\Ha{H${\alpha}$}
\def\Hb{H${\beta}$}
\def\NII{[N\,\textsc{ii}]}
\def\SII{[S\,\textsc{ii}]}
\def\OIIIb{[O\,\textsc{iii}]}
\def\OII{[O\,\textsc{ii}]}

\newcommand{\cgs}{erg\,cm$^{-2}$\,s$^{-1}$}
\newcommand{\ctps}{cnt\,s$^{-1}$}
\newcommand{\kms}{km\,s$^{-1}$}

\title[A TDE in a massive galaxy?]{A tidal disruption flare in a massive galaxy? Implications for the fuelling mechanisms of nuclear black holes}

\author[Merloni et al.]
{\parbox{\textwidth}{A. Merloni$^{1}$\thanks{E-mail: \texttt{am@mpe.mpg.de (MPE)}},
T. Dwelly$^{1}$,
M. Salvato$^{1}$, 
A. Georgakakis$^{1}$,
J. Greiner$^{1}$,
M. Krumpe$^{1}$,
K. Nandra$^{1}$,
G. Ponti$^{1}$,
A. Rau$^{1}$
}\vspace{0.4cm}\\
\parbox{\textwidth}{$^{1}$Max-Planck-Institut f\"ur extraterrestrische
  Physik (MPE), Giessenbachstrasse 1, D-85748, Garching bei M\"unchen,
  Germany\\ 
}}

\begin{document}

\date{}

\maketitle

\label{firstpage}

\vspace{-0.5cm}
\begin{abstract}
We argue that the `changing look' AGN recently reported by LaMassa et
al. could be a luminous flare produced by the tidal disruption
of a super-solar mass star passing just a few gravitational radii
outside the event horizon of a $\sim 10^8 M_{\odot}$ nuclear black
hole. This flare occurred in a massive, star forming galaxy at
redshift $z=0.312$, robustly characterized thanks to repeated
late-time photometric and spectroscopic observations.  By taking
difference-photometry of the well sampled multi-year SDSS Stripe-82
light-curve, we are able to probe the evolution of the nuclear
spectrum over the course of the outburst. The tidal disruption event
(TDE) interpretation is consistent with the very rapid rise and the decay
time of the flare, which displays an evolution consistent with the
well-known $t^{-5/3}$ behaviour (with a clear superimposed re-brightening flare). 
Our analysis places constraints on the physical properties of the TDE, such
as the putative disrupted star's mass and orbital parameters, as well as the
size and temperature of the emitting material.  
The properties of the broad and narrow emission lines observed
in two epochs of SDSS spectra provide further constraints on the
circum-nuclear structure, and could be indicative that the system
hosted a moderate-luminosity AGN as recently as a few $10^4$ years
ago, and is likely undergoing residual accretion as late as ten years after peak, 
as seen from the broad H$\alpha$ emission line. 
We discuss the complex interplay between tidal disruption events and gas accretion
episodes in galactic nuclei, highlighting the implications for future
TDE searches and for estimates of their intrinsic rates.
\end{abstract}

\section{Introduction}
\label{sec:intro}
It is widely accepted that super-massive black holes (SMBH) in the nuclei of galaxies grew over cosmological times mainly by accreting matter from their 
surroundings \citep[][and references therein]{soltan82,merloni15}. Decades of wide and deep surveys at different wavelengths have endowed us with a robust
view of this growth process by constraining the luminosity function of Active Galactic Nuclei (AGN) 
over most of the age of the Universe \citep[see e.g.][and references therein]{hasinger05,hopkins07,aird10,ueda14,buchner15,merloni15,aird15,brandt15}.
Moreover, by studying simultaneously and coherently the properties of the AGN host galaxies within multi-wavelength surveys, 
we can now infer the overall duty-cycle
of the AGN phenomenon, i.e. measure the fraction of time a typical galaxy spends in an active phase above a given (nuclear) luminosity 
\citep{aird12,bongiorno12,hickox14}.
However, the results of these population-based (`snapshot') surveys do not provide any information on the behaviour of accreting black holes on timescales
smaller than those over which galaxies evolve ($\sim$ billions of years). This is due to our uncertain knowledge of the
fuelling mechanisms of AGN: for every galaxy of 
given properties, we still do not know whether black hole growth is rare and long-lived, or frequent and short-lived. 

In fact, understanding how black holes get their fuel at rates sufficient to power the observed AGN population, despite the huge angular momentum
barriers present, is a long-standing goal in the field of galaxy dynamics \citep[see e.g.][and reference therein]{shlosman90,jogee06,hopkins10}.
As summarized in the recent review by \citet{alexander12}, the current consensus is that
gas inflow from kpc scales down to the central $\sim$100\,pc region occurs in all gas-rich galaxies, with the main driving mechanisms
being a variety of externally-excited (mergers, fly-by) or internally-triggered (secular) gravitational instabilities, which help to shed a large fraction 
of  the angular
momentum of the gas. On the other hand, on scales of order 1--10\,pc 
(i.e. close to, or within, the gravitational sphere of influence of the central black hole), 
these large-scale instabilities are less and less effective. Therefore, other dynamical mechanisms must be invoked, such as
dynamical friction, tidal disruption of clouds or star clusters, bars-within-bars, self-gravitating discs, 
eccentric discs or singled-armed spiral modes, 
driven by the complex, time-varying potential of the black hole, stars and gas 
\citep{jogee06,bekki00,shlosman90,hopkins10}.

Such complexity defies any simple modelization, as is apparent in the implementations of black hole
growth within theoretical models of structure formation. There, lacking enough spatial resolution to model 
{\it ab initio} gas and stellar dynamics within galaxies, 
a number of different mechanisms have been postulated as primary drivers of nuclear accretion, often with equally satisfactory results 
\citep[see e.g.][for recent treatments]{hopkins08,hirschmann12,menci14}.

Some observational
attempts to infer recurrence times and durations of AGN accretion episodes 
have been made from indirect (typically space-resolved) tracers of past activity.
A classic example is the light echo of Sgr\,A* that is apparently revealed by refection from the inner molecular 
zone of our own galaxy \citep{sunyaev93,ponti13}. Still within the Galaxy, excess H$\alpha$ emission along the Magellanic Stream
has been claimed to be the result of illumination from Sgr\,A* about $1-3 \times 10^6$ years ago, when the black hole was active at a level of 0.03--0.3 times its Eddington luminosity
\citep{bland13}.
Arguments for powerful past AGN events in nearby galaxies have been based on studies of extended emission-line regions,
which can be used to trace the history of AGN emission over timescales of the order of the
light travel time from the nucleus to the gas \citep[typically $10^4$-$10^5$ years, see, e.g.,][]{dadina10,keel12,gagne14,davies15}.
Radio galaxies offer alternative routes into AGN variability, via the analysis of the morphology of the large-scale radio emission. 
Various pieces of indirect evidence of intermittency have been presented, such as the ripples and
shock waves detected in the X-ray emitting atmosphere around Virgo\,A/M87
\citep{forman05}, or the number versus size counts of small radio galaxies
\citep{reynolds97,czerny09}. Finally, the decrease in the optical depth for Ly$\alpha$
photons due to intervening absorbers along lines of sight to high redshift quasars 
\citep[the so-called `proximity effect'][]{carswell87,lu11}
can also be used to set constraints on the duration of individual QSO episodes.

Yet another complicating factor is that 
at least some part of the fuel supply for AGN must come from stars, which abound within galactic nuclei, and will be tidally disrupted when 
dislodged into orbits passing close enough to the central black hole \citep{hills75,gurzadian81,carter82,rees88,milos06}.
The frequency of such events depends on the stellar dynamical properties of galactic nuclei, and is a non-trivial 
outcome of a series of complex processes \citep{frank76,magorrian99,talexander12,merritt13,vasiliev14}.
Indeed, there is substantial uncertainty in the true rate of tidal stellar disruption in galactic nuclei. 
On the one hand, recent theoretical models, which account for realistic stellar
dynamical models in galactic nuclei of different sizes and masses, have converged towards rates as high 
as $\Gamma_{\rm TDE} \approx 10^{-4}$ yr$^{-1}$, 
with just a weak dependence on the black hole (or host galaxy) mass \citep{wang04,stone15}. 
On the other hand, observational studies have mostly 
reported constraints on the TDE rate that are up to one order of magnitude lower \citep[see, e.g.,][]{donley02,vanvelzen14,khabibullin14}.
This apparent contradiction should not be surprising, as, observationally, the field is still in its infancy; only around two dozen TDE candidates
have been identified so far by means of X-ray 
\citep{komossa_b99,komossa_g99,esquej08,cappelluti09,maksym10,bloom11,burrows11,saxton12,cenko12,khabibullin14}, 
UV  \citep{gezari08,gezari09} and optical \citep{vanvelzen11,gezari12,arcavi14,chornock14,holoien14} observations 
\citep[see][for a recent overview]{gezari14}. 
In particular, the full extent of the various selection effects that may plague 
TDE selection in different bands of the electromagnetic spectrum have yet to be properly characterized.
Nevertheless, the rate of discovery is increasing dramatically with the advent of large wide-area optical time-domain surveys 
(mostly driven by supernovae searches), 
such as the Catalina Real-Time Transient Survey \citep[CRTS,][]{drake09}, the Palomar Transient Factory \citep[PTF,][]{rau09}, 
PanSTARRS \citep{kaiser10} and the All-Sky Automated Survey for Supernovae \citep[ASA-SN,][]{shappee14}. 
The future promises even more rapid advances with next generation wide area X-ray \citep[SRG/eROSITA,][]{merloni12,khabibullin14a} 
and optical (Skymapper, \citealt{keller07}, ZTF\footnote{{\tt http://www.ptf.caltech.edu/ztf}}, and LSST, \citealt{ivezic08}) surveys. 

In this work, we focus on just one particular example: a $z=0.312$ galaxy within the Sloan Digital Sky Survey (SDSS) `Stripe-82' 
area\footnote{Stripe-82 is a $109 \times 2.5$ degree survey field that lies along the celestial 
Equator within the Southern Galactic Cap, which has been imaged multiple times by the SDSS camera} which has undergone
a dramatic and rapid change in its nuclear emission, from type-1 (broad line) AGN-like, to type-1.9 
(showing only a weak broad H$\alpha$ emission line) within a 10 year interval.
\citet{lamassa15}, who first reported this serendipitous discovery, rightly point out that such a dramatic change is rare
\citep{dadina10,denney14}, and can provide 
important clues on the nature of black hole fuelling in galactic nuclei.
Here we perform an in-depth analysis of the available (mostly public) data on this source (section~\ref{sec:data} and \ref{sec:bol_lc}), 
and we argue, 
based on the outcome of our analysis, that its overall light-curve in the last decades is consistent with the tidal disruption
of a main-sequence star (with mass $M_*>1 M_{\odot}$) by a $\approx 10^8 M_{\odot}$ black hole hosted in the nucleus of a massive, star-forming galaxy
(section~\ref{sec:tde}). Our conclusions are mainly based on the properties of the optical light-curve, and the overall energetics of the event.
However, the full picture reveals the complexity of this galactic nucleus, as we demonstrate that (i) the material giving rise to the 
broad emission lines illuminated by the observed flare cannot be a distant part of the stellar debris (section~\ref{sec:blr}), and (ii)
the narrow emission lines could indicate that about $10^4$ years ago the central black hole was active at a level similar to that observed  
(section~\ref{sec:nlr}). This prompted us, in section~\ref{sec:rates}, to develop a simple unified model for AGN fuelling, in order to estimate
the relative occurrence of gaseous accretion and tidal disruption flares in nearby galactic nuclei. 
Finally, we draw our conclusions in section~\ref{sec:conclusions}.

\section{Data analysis}
\label{sec:data}
\subsection{Source identification and long term evolution}
\label{sec:lc}
The main focus of this paper is the object SDSS~J015957.64+003310.5,
hereafter SDSS~J0159+0033.  SDSS~J0159+0033 first came to our
attention (before being independently reported by \citealt{lamassa15})
whilst creating a reference sample of bright X-ray sources to guide
target selection for the SDSS-IV/SPIDERS survey.
(Merloni et al., in prep., see also {\tt www.sdss.org/surveys/}).
Starting from a parent sample of bright
($F_{\rm 0.2-2 keV}>10^{-13}$\,erg\,cm$^{-2}$\,s$^{-1}$) X-ray sources
taken from the 3XMM serendipitous source catalog\footnote{{\tt
    http://xmmssc-www.star.le.ac.uk}}, we found 557 objects having
optical spectroscopic observations in the latest SDSS data release (DR12, \citealt{alam15}). 
A small subset (35/557) of these had a
good quality spectrum available from both the SDSS-I/II phase 
\citep[using the original SDSS spectrograph,][]{york00}, as well as a spectrum obtained during
the SDSS-III phase \citep[using the upgraded BOSS spectrograph,][]{smee13}. Within this
subset of 35 objects there was a single one, SDSS~J0159+0033, for which
the automated classification derived from the SDSS-I/II spectrum
({\tt CLASS=`QSO'}) disagreed with that derived from the SDSS-III
spectrum ({\tt CLASS=`GALAXY'}). 
SDSS~J0159+0033 appeared in our original X-ray sample due to its serendipitous 
detection in an {\it XMM-Newton} observation of the nearby luminous QSO Mkn\,1014 
\citep{markarian77}.
The apparent disparity between the two spectroscopic classifications of this 
object prompted further investigation.



SDSS~J0159+0033 was first imaged by the SDSS camera on 1998 September
25, and was subsequently selected for spectroscopic follow up within the
low redshift {\it ugri} colour selected QSO target category
(specifically with target bit {\tt TARGET\_QSO\_SKIRT},
\citealt{richards02,stoughton02}). The pipeline measurements derived
from the SDSS photometry found SDSS~J0159+0033 to be a marginally
resolved galaxy having de-reddened {\it ugri} colours ($u-g=0.53$,
$g-r=0.79$, $r-i=0.31$) that placed it outside the main stellar locus,
blueward of the colour regions used to reject `normal' galaxies, and
also outside additional colour boxes designed to reject rarer stellar
contaminants.
The optical spectrum taken on 2000 November 23 (MJD 51871) is classified as a $z=0.312$ broad line QSO, due the unmistakable 
presence of a blue continuum and 
broad Balmer lines (H$\alpha$ and H$\beta$, see section~\ref{sec:spectra} for more details).

On 2010 January 5 (MJD 55201), SDSS~J0159+0033 was observed spectroscopically 
by the Baryon Acoustic Oscillations Survey \citep[BOSS,][]{dawson13} within the SDSS-III project, as part of a small
random sub-set of known broad line AGN from SDSS that were re-observed in order to
define QSO spectral templates for BOSS (targeting flag {\tt
 TEMPLATE\_QSO\_SDSS}).  However, the resulting
spectrum revealed a $z=0.312$ star-forming galaxy with essentially no
blue continuum, and no prominent broad emission lines, apart from a
weak broad shoulder to the H$\alpha$ line. The BOSS pipeline measured
a velocity dispersion of $\sigma=169 \pm 24$ km/s.

Five X-ray observations of the field containing SDSS~J0159+0033 exist in public archives. 
The oldest of these is a Roentgen Satellite (ROSAT) All-Sky-Survey
scan from 1991 January, which failed to detect the source, placing a 3$\sigma$ upper limit of $3.2 \times 10^{-2}$\,\ctps, corresponding
to an unabsorbed (i.e. corrected for the Galactic column density of $2.3 \times 10^{20}$\,cm$^{-2}$, \citealt{kalberla05})
X-ray flux limit in the 0.1--2.4 keV band  $F_{\rm 0.1-2.4 keV}< 8 \times 10^{-13}$\,\cgs\ 
(for a $\Gamma_X=2.1$ power-law spectrum).
Two deeper (6.1 and 1.6 ksec), {\it ROSAT-PSPC} pointed observations of nearby sources 
(Mkn\,1014 and MS\,0158.5+0019) on 1992 January 17 and 1992 July 24 also covered the location of SDSS~J0159+0033,
placing more stringent upper limits on its flux of 
$F_{\rm 0.1-2.4 keV}< 2.6 \times 10^{-14}$ and $F_{\rm 0.1-2.4 keV}< 4.6 \times 10^{-13}$\,\cgs\ respectively.
In contrast, the {\it XMM-Newton} 
observations of the same region performed in 2000 July (aimed at the nearby luminous QSO Mrk\,1014), 
caught SDSS~J0159+0033 in an X-ray bright state, with a flux of 
$F_{\rm 2-10 keV}=2.6 \times 10^{-13}$\,\cgs, and a spectral shape typical of accreting
super-massive black holes, characterized by a power-law with slope $\Gamma_X=2.1$, and
no sign of significant absorption at the redshift of the source\footnote{Within this model, this 
corresponds to $F_{\rm 0.1-2.4 keV} = 6.4 \times 10^{-13}$\, \cgs.} \citep{lamassa15}.
Finally, a {\it Chandra} observation of Mkn\,1014 serendipitously covered SDSS~J0159+0033 in 2005, 
and measured it to be a factor of $\sim$7 fainter than in 2000, but again
with a power-law spectrum consistent with no absorption at the redshift of the source.

\begin{figure*}
\includegraphics[width=0.65\textwidth,angle=270,clip]{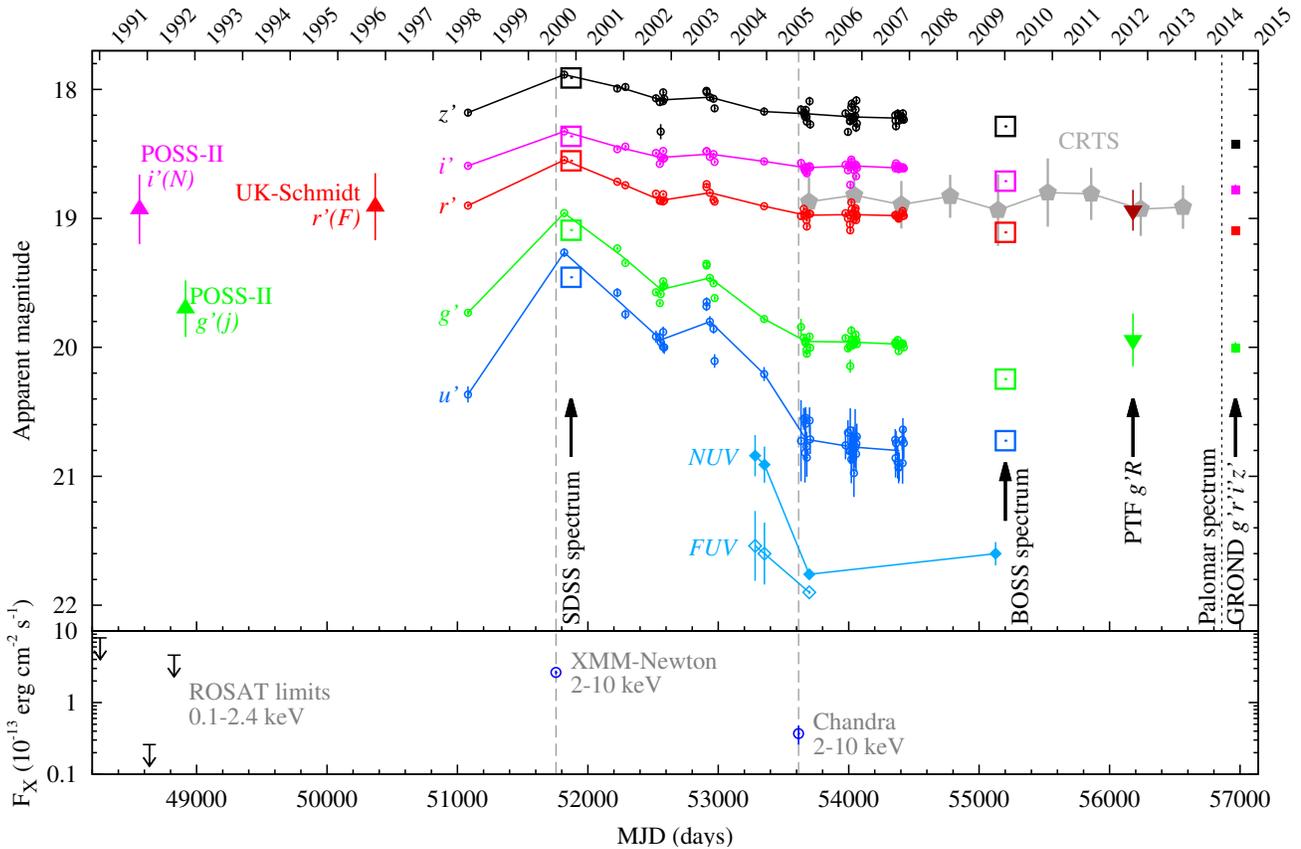}
\caption{{\it Top Panel}: Long term optical light-curve for SDSS~J0159+0033. The SDSS
  Stripe-82 photometric measurements in the {\it u'g'r'i'z'} filters
  are shown with small open circle symbols (blue, green, red, magenta
  and black respectively), with the yearly medians for each filter
  connected by solid lines. Stripe-82 epochs not meeting the
  data-quality criteria described in the text are omitted.  The
  spectro-photometric measurements derived from the two epochs of SDSS
  spectroscopy (also for the {\it u'g'r'i'z'} filters) are shown with
  large open boxes (offset by $-0.3$ magnitudes to align roughly with
  the contemporary SDSS photometric measurements). Older photometric
  measurements derived from the Palomar and UK-Schmidt photographic
  plate surveys, are individually labelled \citep[GSC2.3.2][]{lasker08}. They 
have been converted to the nearest SDSS filters according to the procedure 
described in Appendix ~\ref{appendix_dss}. 
The yearly median and RMS of the
  unfiltered CRTS (Catalina Transient Survey, \citealt{drake09})
  measurements are shown with (grey) filled pentagons. The median and
  RMS of the public PTF (Palomar Transient Factory, \citealt{rau09})
  data in the {\it g'} and {\it R} filters are shown with
  downward-pointing triangles (green and red respectively). GROND
  photometry \citep{greiner08} in the {\it g'r'i'z'} bands is shown
  with filled square symbols (green, red, magenta and black
  respectively). {\it GALEX} \citep{martin05} data-points are shown with filled and
  open light blue diamonds for the {\it NUV} and {\it FUV} bands
  respectively.  The epochs of the {\it ROSAT}, {\it XMM-Newton} and
  {\it Chandra} X-ray observations are shown with vertical dashed
  lines. The {\it Bottom panel}: shows the X-ray lightcurve.}
\label{fig:long_term_lc}
\end{figure*}

To place the observed high-amplitude variability, 
detected in both X-rays and SDSS optical spectra, into a long-term context, 
we have collected publicly available UV and optical photometric data-points
of the source spanning the last $\sim$30 years, which we display in Fig.~\ref{fig:long_term_lc}.

There is no evidence for significant photometric variability after 2005. 
The spectral energy distribution (SED) of SDSS~J0159+0033 derived from the 
most recent Stripe-82 photometric observations (autumn 2007) shows a great deal of consistency with the 
BOSS spectrum taken in 2010 (see Fig.~\ref{fig:spectrum_lc_left}, and section~\ref{sec:spectra} below). 
As a further check, we obtained 7-band ({\it g'r'i'z'JHK}) GROND \citep{greiner08} observations of SDSS~J0159+0033 on 
2014 November 04, and found that the SED was still dominated by the host galaxy, fully consistent with all photometric observations since 
about 2005,   
including public photometry from wide-field, high-cadence, optical transient surveys (CTRS and PTF),
and also confirmed by the Palomar spectrum taken in 2014 (presented by \citealt{lamassa15}).

More interestingly, we find that also all available data-points 
before 1998 are consistent 
with the optical emission of the source being dominated by its host galaxy, 
albeit with larger uncertainties due to the poorer sensitivity and calibration of photographic plates. 
The older photometric measurements shown in fig.~\ref{fig:long_term_lc} were derived from the Palomar and UK-Schmidt photographic
  plate surveys \citep[GSC2.3.2,][]{lasker08}, and 
have been converted to the nearest SDSS filters according to the procedure 
described in Appendix ~\ref{appendix_dss}. These also include  a 1983 V-band observation ("Palomar Quick-V
Northern Survey", not shown in Fig.~\ref{fig:long_term_lc}) which translates to {\it r'}$=19.17\pm 0.36$ 
(see Tab.~\ref{tab:photo} in Appendix~\ref{appendix_dss}).
Uncertainties on those fluxes, however, 
are such that it is not possible to exclude that SDSS~J0159+0033 was in a state of (low-level) AGN activity even before the flare began.

In summary, the only period during the last $\sim$30 
years when SDSS~J0159+0033 was significantly brighter than 
its late-time (post-2005) constant level (both in X-rays and optical bands) was during the period between 1998 and 2005, 
suggesting a one-off, high amplitude, flaring episode,
rather than a more typical AGN variability pattern, often 
modelled as red-noise or ``damped random walk'' \citep[][and references therein]{macleod10}. 

\subsection{Spectral analysis}
\label{sec:spectra}
\citet{lamassa15} have presented a detailed analysis 
of the 2000 and 2010 spectra, and derived the mass of the central black hole under the assumption that the observed broad H$\alpha$ and H$\beta$
emission lines are produced by virialized material in orbit around the central black hole. Adopting the relation between observed line width,
continuum luminosity and black hole mass of \citet{vestergaard06} and \citet{greene10} for H$\beta$  and H$\alpha$, respectively, they derived
a mass of about $1.7 \times 10^8 M_{\odot}$. Here we briefly describe the results of our independent, detailed, spectral analysis.

\begin{figure}
\includegraphics[width=0.5\textwidth,clip]{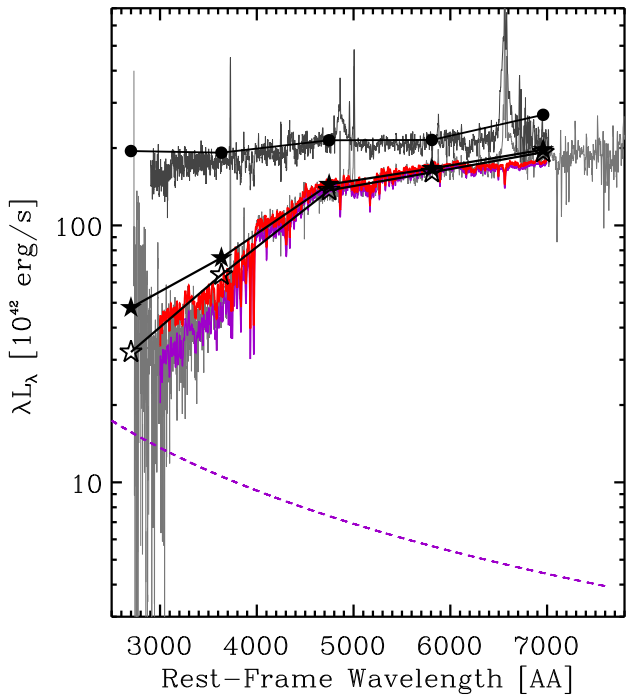}
\caption{The two dark grey lines show the calibrated spectra ($\lambda L_{\lambda}$, in units of $10^{42}$ erg/s) 
of SDSS~J0159+0033 from 2000 (upper curve) and from 2010 (lower curve).
Superimposed are the Stripe-82 {\it u'g'r'i'z'} photometric data points from near the outburst peak (2000, black solid circles), and 
the weighted average of the last season of Stripe-82 photometry (2007, black solid stars). 
The red solid line is the best-fit model to the continuum emission of the 2010 spectrum, which as described in detail in the text, 
is the sum of a SSP host galaxy spectrum (solid purple line) plus a power-law continuum, whose slope has been fixed at 
$F_{\rm lambda}\propto \lambda^{-2.33}$ (purple dashed line). 
Finally, the open star symbols show the implied `pure' galaxy SED that we use to study the nuclear flare evolution. }
\label{fig:spectrum_lc_left}
\end{figure}

The SDSS and BOSS spectra were obtained through different sized apertures (3.0 and 2.0 arcsec diameter respectively). 
Therefore we first corrected the spectro-photometric calibration of the SDSS (2000) and BOSS (2010) spectra by matching the 
{\it i'}-band spectro-photometric flux measurements
(which, at the redshift of the source, $z=$0.312, is devoid of strong emission lines), to the nearest (in time) 
Stripe-82 {\it i'}-band photometric
measurements. For the SDSS 2000 spectrum this was the single Stripe-82 data-point closest to the peak of the outburst, 
and for the BOSS 2010 spectrum we 
took a weighted average over the autumn 2007 {\it i'}-band photometric measurements (see figure~\ref{fig:spectrum_lc_left}).

We then fitted the BOSS 2010 spectrum in the regions of the most prominent emission lines 
(OII, $3650{\rm \AA}<\lambda<3800{\rm \AA}$ ; H$\beta$, $4700{\rm \AA}<\lambda<5100{\rm \AA}$; 
H$\alpha$, $6480{\rm \AA}<\lambda<6750{\rm \AA}$). 
For the host galaxy continuum and absorption 
lines, we used the SSP (single stellar population) high-resolution evolutionary model templates from \citet{gonzalez05}, 
the preferred model (from a comparison with the broad band SED) being a $Z=0.019$ (i.e. $\sim$ solar metallicity) `Padova' isochrone
template with an age of 2.5 Gyr. 
The normalization of this component corresponds to a total mass of formed stars of $\approx 1.1  \times 10^{11} M_{\odot}$, 
which, taking into account stellar mass loss, would imply a stellar mass of the host galaxy of about $\approx 8\times 10^{10} M_{\odot}$\footnote{We note that we are not aiming at a precise determination of the age of the stellar population. Indeed, if we chose a younger SSP from the same library
(for example a 1.6 Gyr old one, in agreement with LaMassa et al. 2015), we would recover an equally good fit to the narrow wavelength ranges around the 
emission lines we are interested in, with all emission line parameters within the errors, and a stellar mass of the host of 
$\approx 7\times 10^{10} M_{\odot}$. This shows that, on the other hand, the total stellar mass of the system is quite robustly 
determined by the SSP normalization.}.
The corresponding model SED is shown 
in Fig.~\ref{fig:spectrum_lc_left}, and 
in the bottom panels of Fig.~\ref{fig:broad}. To the stellar component, we added 
a power-law continuum to represent any residual nuclear emission, which is required to fit the blue end of the observed spectrum. However, the slope of such a component is 
barely constrained by the BOSS 2010 spectrum, 
and so we decided to fix it to that expected for the `canonical' viscous accretion disc \citep{ss73} $\alpha=-1/3$, where $F_{\lambda} \propto \lambda^{-2+\alpha}$. Changing the slope of the power-law component 
within the range observed from the flare emission during the outburst (see section~\ref{sec:bol_lc} below) 
changes our results only marginally, and not qualitatively. 

We fit the emission lines with the {\tt MPFIT} IDL routine \citep{markwardt09}, using a set of Gaussians, as described below; 
the best-fit parameters for the emission lines (of both the SDSS 2000 and BOSS 2010 spectra) are shown in Table~\ref{tab:lines}. 

For the \Hb\ region in the BOSS 2010 spectrum (see top left plot in fig.~\ref{fig:broad}), one narrow component is sufficient to fit
the Balmer line, while the \OIIIb\ complex (where we fixed the flux of \OIIIb~4959\,${\rm \AA}$ to be $1/3$ of the \OIIIb~5007\,${\rm \AA}$ flux) requires an additional, blue-shifted
broad component, typically associated with outflows in the narrow-line region ionized gas. This blue wing has a measured FWHM of 774$\pm$87 \kms\ and is blue-shifted
by about 100$\pm$50 \kms\ with respect to the narrow component. Such a width for the broad component of \OIIIb\ is well within the average for Seyfert 2 galaxies in SDSS of similar luminosity as SDSS~J0159+0033 \citep{mullaney13}, but not so high as to place it squarely among the secure sources of AGN-driven outflows (typically requiring
FWHM $> 1000$ \kms, see e.g. \citealt{brusa15}).

In the \Ha\ region of the BOSS 2010 spectrum (see top right plot in fig.~\ref{fig:broad}), in addition to the narrow \Ha\ line and \NII\ complex 
(where, again, we fixed the flux of \NII~6548 ${\rm \AA}$ to be $1/3$ of the \NII~6583 ${\rm \AA}$ flux), 
a very broad line is required, 
as already noticed by \citet{lamassa15}.

\begin{figure}
\includegraphics[width=0.5\textwidth,clip]{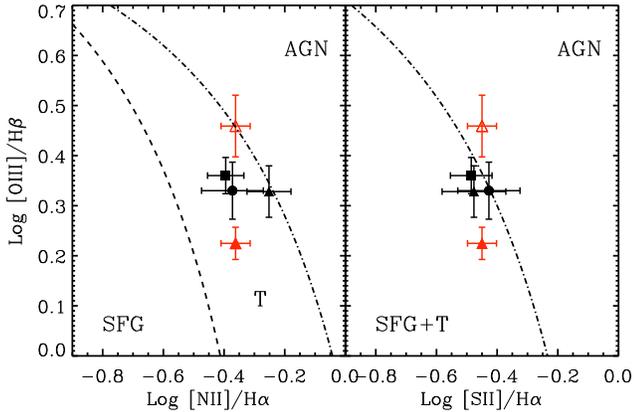}
\caption{Line emission diagnostic (BPT, \citealt{bpt}) diagrams for the three spectroscopic observations (2000 SDSS, black circle; 2010 BOSS, triangle and 2014, square) of SDSS~J0159+0033 presented in \citet{lamassa15}. The red triangles show the results of our own analysis of the 2010 spectrum: the filled symbol corresponds to the case in which we
only consider the narrow \OIIIb\ line component, the empty one for the sum of broad plus narrow \OIIIb\ line components. 
In each panel the dot-dashed line represents the \citet{kewley06} separation between
pure AGN and the rest of the galaxy population. In the left panel the dashed line is the separation between star-forming galaxies and `transition' objects (i.e. those with ionizing photons coming from both star-forming regions and AGN) defined by \citet{kauffmann03}.}
\label{fig:bpt}
\end{figure}

\begin{figure*}
\includegraphics[width=0.48\textwidth,clip]{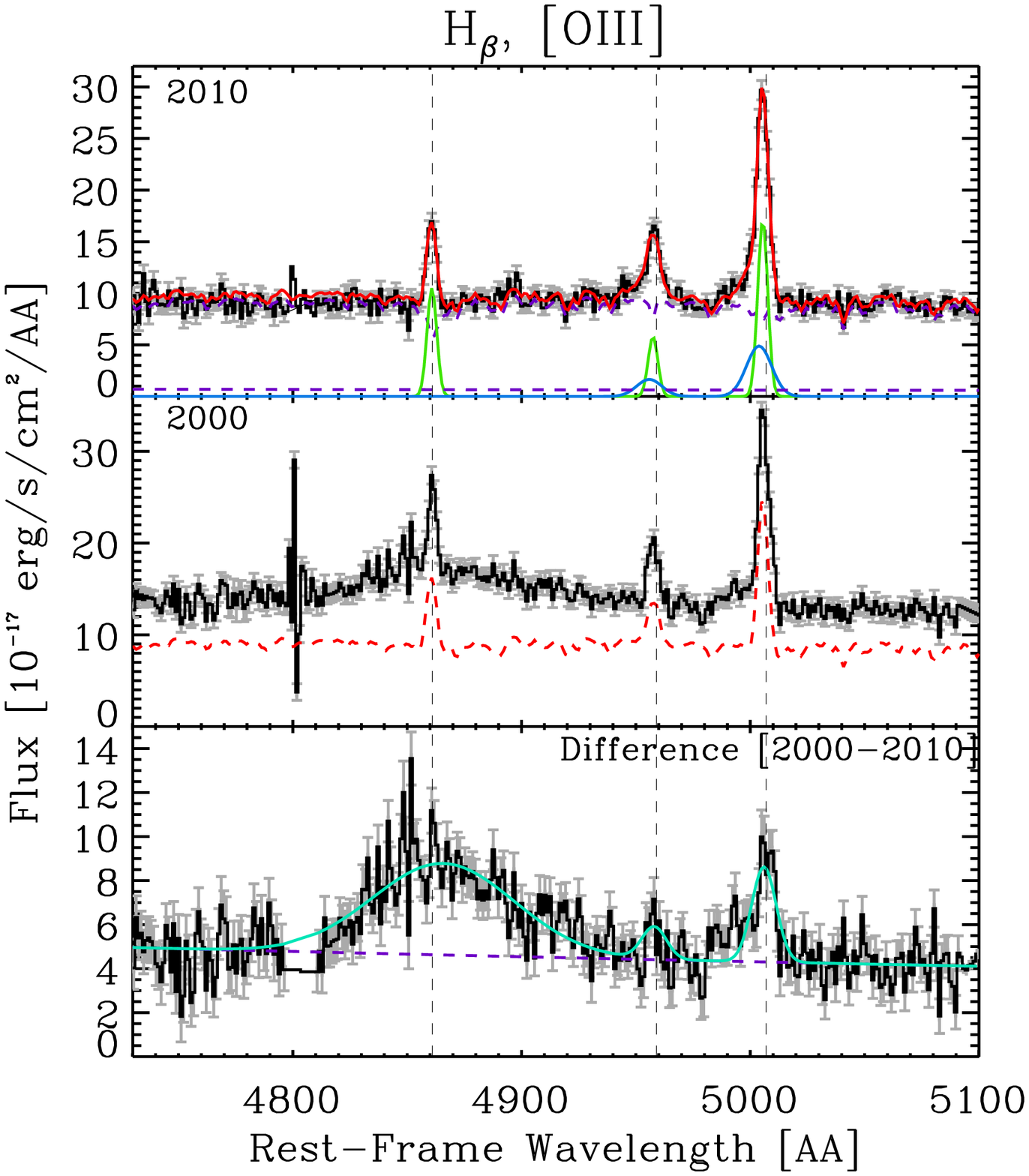}
\includegraphics[width=0.48\textwidth,clip]{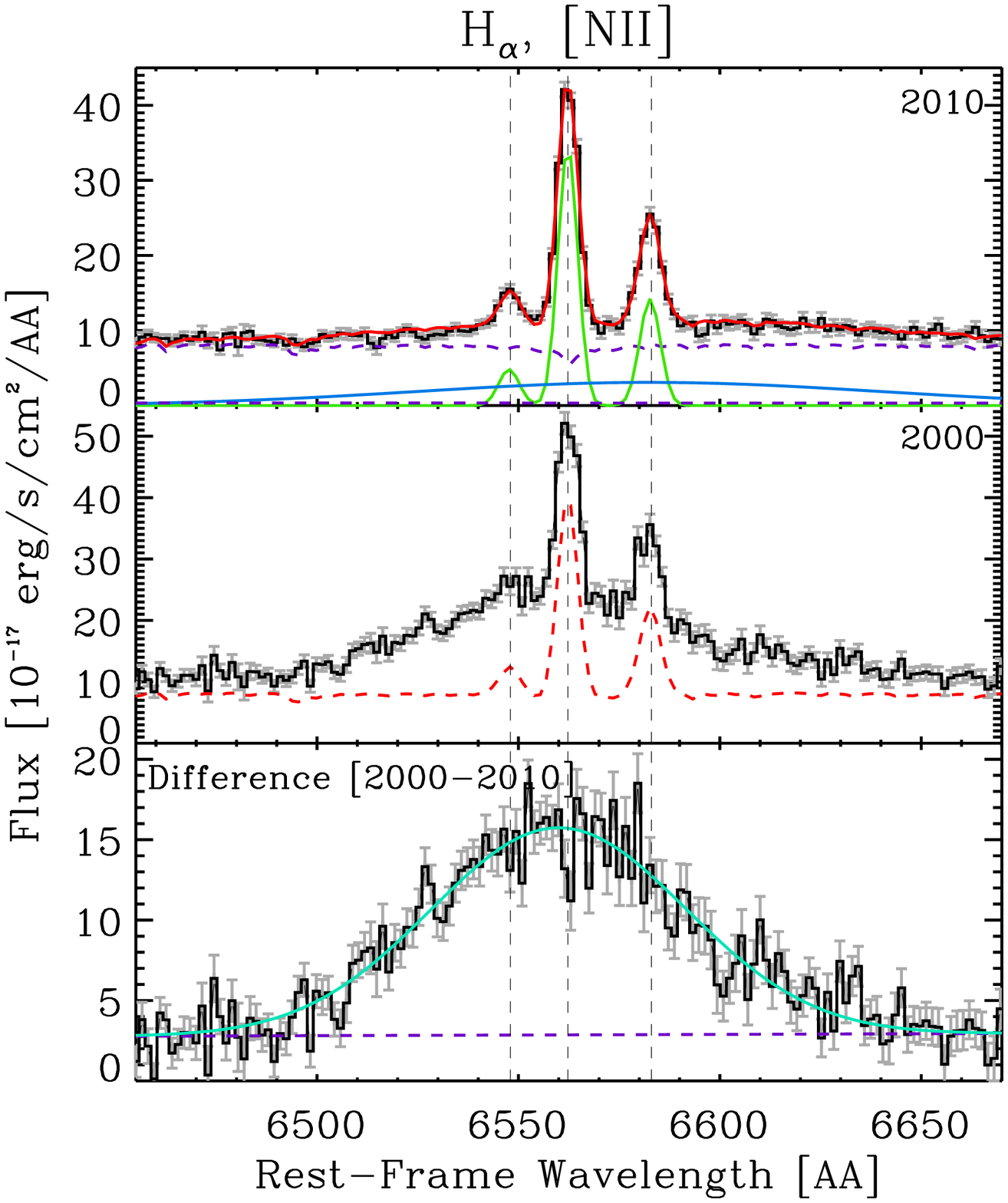}
\caption{{\it Left:} Spectral analysis of the \Hb\ region. The top panel shows the calibrated BOSS 2010 spectrum 
with uncertainties 
(black histogram with grey error bars), fitted with a host galaxy continuum model from Gonzalez Delgado et al. (2005) 
(a 2.5 Gyrs old $Z=0.019$ SSP; top purple dashed line), a power-law continuum of slope $\alpha=-1/3$ (bottom purple dashed line), 
narrow H$\beta$ and \OIIIb\ emission lines (green solid line) and a broad \OIIIb\ component (blue solid line). 
The middle panel shows the observed, calibrated 2000 SDSS spectrum, together with the 2010 best fit host galaxy model 
continuum plus narrow emission lines (red dashed line).
The bottom panel  shows the difference spectrum (2000--2010), where the rapidly time-varying emission components emerge. This difference spectrum is fitted with a
power-law of slope $\alpha=-0.5$, where $F_{\lambda} \propto \lambda^{-2+\alpha}$ (dashed purple line) plus broad H$\beta$ and \OIIIb\ emission lines (cyan line).
{\it Right:} Spectral analysis of the \Ha\ region. The top panel shows the calibrated BOSS 2010 spectrum 
with uncertainties 
(black histogram with grey error bars), fitted with a host galaxy continuum model from Gonzalez Delgado et al. (2005) 
(a 2.5 Gyrs old $Z=0.019$ SSP; purple dashed line), a power-law continuum of slope $\alpha=-1/3$ (bottom purple dashed line), 
a narrow H$\alpha$ and [NII] emission lines (green solid line) and a broad H$\alpha$ component (blue solid line). 
The middle panel shows the observed, calibrated 2000 SDSS spectrum, together with the 2010 best fit model host galaxy 
continuum plus narrow emission lines (red dashed line).
The bottom panel finally shows the difference spectrum (2000--2010), where the rapidly time-varying emission components emerge. This difference spectrum is fitted with a
power-law continuum (dashed purple line) plus broad H$\alpha$ emission line (cyan line).}
\label{fig:broad}
\end{figure*}

The narrow lines observed in the BOSS 2010 spectrum can be used to investigate the source of ionizing radiation that excited them.
We plot in Fig.~\ref{fig:bpt} the BPT \citep{bpt} diagnostic diagram for our measurements of the narrow emission lines (\OIIIb\, \Hb, \NII\ and \Ha) 
as well as those derived 
from the analysis of the three spectra presented by \citet{lamassa15} (the 2000 SDSS and 2010 BOSS spectra, as well as a later Palomar DBSP spectrum taken in 2014). 
As expected, since we fit two components to the \OIIIb\ lines, we find different H$\beta$/\OIIIb\ flux ratios to those of 
\citet{lamassa15}, who adopt a single component model,
whilst the \NII/\Ha\ and \SII/\Ha\ line ratio measurements are in agreement.
If we only consider the narrow components of \OIIIb, SDSS~J0159+0033 appears to lie well within the `transition' region between star-forming galaxies and AGN. On the other hand, 
if we sum together both narrow and broad components of the \OIIIb\ line, the object moves into the AGN-dominated part of the diagram. Thus, it 
appears that the line emitting material was probably ionized by a source harder than
that associated with pure star-formation. AGN activity would of course provide a sufficiently hard ionizing source, 
but the exact amount of any putative AGN contribution to the observed narrow 
line emission is harder to disentangle. We discuss the consequence of the narrow emission lines properties for the 
interpretation of the flare in section~\ref{sec:nlr}.

Moreover, the strong [OII] emission line can also be used to infer an approximate star formation rate (SFR) of the host. We follow the 
procedure of \citet{silverman09} to correct the [OII] for any possible AGN component, by assuming that all observed [OIII] emission is AGN-driven, thus obtaining 
a lower limit to the derived SFR. We assume a fixed [OII]/[OIII]$=0.21$ ratio for AGN-only excited emission lines, and use the difference $L_{\rm [OII]}-L_{\rm [OII],AGN}$ to derive a SFR \citep[see eq. (2) of][]{silverman09}. Under this assumption, we obtain log SFR$\approx 1.6 (M_{\odot}$/yr).

A  direct comparison between the emission lines seen in the 2010 BOSS and 2000 SDSS spectra can reveal the variable components, and help 
determine their physical origin.
In our analysis of the 2000 SDSS spectrum, taken very close in time to the photometric peak of the light-curve (see Fig.~\ref{fig:long_term_lc}), 
we have first assumed that the host galaxy continuum and the narrow emission lines seen in the 2010 BOSS spectrum have not changed significantly over the intervening ten years. 
This baseline constant spectrum is shown as a dashed red line in the middle panels of fig.~\ref{fig:broad}. We then take the difference
of the two spectra (difference = 2000 SDSS spectrum -- 2010 host galaxy emission model), plotted in the bottom panels of Fig.~\ref{fig:broad}. In the \Hb\ region, a clear rising continuum is visible, with a prominent broad
\Hb\ emission line. We fix the power-law continuum slope to that measured over the entire spectral range in the difference spectrum, $\alpha=-0.5$, leaving its normalization free, and measure the flux and FWHM of the broad lines (also reported in table~\ref{tab:lines}). The \OIIIb\ emission region is noisy, but the spectrum
taken at the peak of the outburst does also show a broad component, which is consistent with having the same properties of that observed ten years later 
(FWHM of 753$\pm$80 km/s).

We can use the measured parameters of the broad emission lines to estimate the mass of the central black hole, using the standard `single epoch' virial method 
\citep[see e.g.][and references therein]{peterson04}. We adopt the \citet{Greene07} scaling between black hole mass, \Ha\ FWHM and \Ha\ line luminosity 
to derive log$(M_{\rm BH}/M_{\odot}) \simeq 7.8$ and log$(M_{\rm BH}/M_{\odot}) \simeq 8.2$ for the 2000 and 2010 spectra, respectively. Scaling laws to 
obtain black hole masses from
the measured width of the H$\beta$ line and the continuum luminosity at 5100\,${\rm \AA}$ (in 2000 measured to be $\lambda L_{5100} = 6.8 \times 10^{43}$\,erg\,s$^{-1}$) 
have been published by many authors. Adopting the calibration of 
\citet{vestergaard06}, we derive log$(M_{\rm BH}/M_{\odot}) \simeq 8.1$, 
while following \citet{greene10} we get log$(M_{\rm BH}/M_{\odot}) \simeq 8.2$. 
All these estimates come 
with both statistical and systematic uncertainties of about 0.3 dex, due to the uncertain calibration and the unknown geometry of the broad line region itself 
\citep[see e.g.][]{shen12}. Encouraged by the consistency among all the virial black hole mass estimators \citep[see also][]{lamassa15},
we consider these estimates robust, and, for the remaining of the paper, assume a fiducial SMBH mass of $10^8 M_{\odot}$.

\begin{table*}
\caption{Emission line parameters derived from from our spectral analysis (see section \ref{sec:spectra} for details).}
\label{tab:lines}
\begin{tabular}{lcccccc}
\hline
 &\multicolumn{3}{c}{SDSS 2000} & \multicolumn{3}{c}{BOSS 2010}\\
 & $\lambda ({\rm \AA})$ & FWHM (km\,s$^{-1}$) & $L_{\rm line} (10^{40} {\rm erg\,s^{-1}})$ & $\lambda ({\rm \AA})$ & FWHM (km\,s$^{-1}$) & $L_{\rm line} (10^{40} {\rm erg\,s^{-1}})$ \\
\hline
$[{\rm OII}]$ & 3726.79$^a$ & 468$^a$ & 38.7$^a$ & 3726.79$\pm$0.07 & 468$\pm$13 & 38.7$\pm$3.0 \\
H$\beta_{\rm narrow}$ & 4860.8$^a$ & 309$^a$ & 18.0$^a$ & 4860.8$\pm$0.1 & 309$\pm$13 & 18.0$\pm$1.8 \\
H$\beta_{\rm broad}$ & 4865.8$\pm$1.4 & 4493$\pm$234 & 103.9$\pm$5.1 & -  & - & - \\
$[{\rm OIII}]_{\rm narrow}$ & 5005.7$^a$ & 308$^a$ & 30.2$^a$ & 5005.5$\pm$0.1 & 308$\pm$13 & 30.2$\pm$2.2 \\
$[{\rm OIII}]_{\rm broad}$ & 5006.0$\pm$0.5 & 753$\pm$80 & 18.6$\pm$2.6 & 5003.8$\pm$0.6 & 774$\pm$87 & 21.6$\pm$5.1 \\
H$\alpha_{\rm narrow}$ & 6562.26$^a$ & 262.2$^a$ & 67.9$^a$ & 6562.26$\pm$0.05 & 262.2$\pm$5.1 & 67.9$\pm$4.1 \\
H$\alpha_{\rm broad}$ & 6559.9$\pm$0.9 & 3408$\pm$110 & 329$\pm$11 & 6582.4$\pm$2.0  & 6167$\pm$280 & 143.0$\pm$6.7 \\
$[{\rm NII}]$ & 6582.7$^a$ & 276$^a$ & 29.5$^a$ & 6582.7$\pm$0.1 & 276$\pm$13 & 29.5$\pm$2.9\\
$[{\rm SII}]$ & 6715.8$^a$ & 358$^a$ & 24.1$^a$ & 6715.8$\pm$0.2 & 358$\pm$18 & 24.1$\pm$2.2\\
\hline
$^a$Line parameter fixed at the 2010 values.
\end{tabular}
\end{table*}

In sections 
\ref{sec:blr} and \ref{sec:nlr} we discuss the observed properties of the emission line regions within the context of our interpretation of the observed flare from SDSS~J0159+0033.

\section{The bolometric light-curve of the flare}
\label{sec:bol_lc}
A change in the optical spectral properties of AGN as dramatic as that observed in SDSS~J0159+0033, 
and taking place within just ten years, is very unusual.
\citet{lamassa15} have discussed a possible interpretation of these rapid changes of flux and spectral properties 
within the context of the extreme end of normal AGN variability.
Given the implied size of the broad line region, and the observed X-ray spectra (both {\it XMM-Newton} and {\it Chandra} can be
fitted with un-absorbed power-laws), they conclude that an obscuration event, in the form of a large-scale cloud passing in front of the
central source and of the broad line region, is not consistent with the timescale of the observed fast flare, nor with
the spectral slope of the 2005 {\it Chandra} spectrum. It appears, then, that the  intrinsic emission of the AGN is fading rapidly, 
possibly due to a decrease of the accretion rate onto the central black hole.
However, in the context of standard accretion-disc theory, it is not straightforward to explain both the rapid fading in 
the period 2000--2005 and, in particular, the very fast brightening of the source between 1998 and 2000; viscous timescales of optically 
thick and geometrically thin accretion discs as close as 10 gravitational radii from a 
$10^8 M_{\odot}$ black hole, with the luminosity observed close to its 
peak, are of the order of a few years (see section~\ref{sec:discussion} below). Larger discs, needed if the observed optical light is produced from viscous
dissipation within the disc itself, would evolve on even longer timescales.

It appears, then, that a physical process is required that can give rise to rapid enhancement of the accretion rate,
a process that delivers large quantities of matter very close to the central black hole, where viscous (or thermal) 
times are the shortest. Moreover, such 
a process must be be highly intermittent, because, as we discussed before, 
the long term light-curve of SDSS~J0159+0033 is essentially dominated by this 
one single flare. In the following, we describe why we believe that a Tidal Disruption Event (TDE) 
is a much more natural interpretation for this peculiar transient event.

We first analyze in more detail the evolution of the flare, as traced by the Stripe-82 lightcurve. 
Our starting point is the Stripe-82 photometry
for SDSS~J0159+0033 and two neighbouring reference stars (SDSS\,J020002.27+003250.6, a $r\approx 18$ magnitude star about 1.2 arcmin away, and SDSS \,J015955.54+003419.6, a $r\approx 16.1$ star about 1.26 arcmin away),
retrieved from the SDSS DR7 data server\footnote{{\tt http://cas.sdss.org/stripe82/en/tools/crossid/crossid.asp}}.
In the few cases where multiple data-points were found with the same MJD, we took daily averages, resulting in 70
independent epochs of photometry. We then cleaned the light curve by excluding all 
epochs with poor photometry, as deduced by comparing the flux of each reference star in each epoch 
and each band with their long-term average, and discarding all epochs where the average
flux of the two stars in any band is more than 2$\sigma$ off its mean. This leaves us with 50 high quality
epochs of photometry. 


\begin{table*}
\caption{Nuclear (host-subtracted) light-curve of SDSS~J0159+0033 in the {\it u'g'r'i'z'} SDSS bands}
\label{table:lc}
\begin{tabular}{lccccc}
\hline
MJD & $L_{\lambda,u'}\Delta \lambda_{u'}$ & $L_{\lambda,g'}\Delta \lambda_{g'}$ & $L_{\lambda,r'}\Delta \lambda_{r'}$ & $L_{\lambda,i'}\Delta \lambda_{i'}$ & $L_{\lambda,z'}\Delta \lambda_{z'}$ \\
    & $(10^{42} {\rm erg/s})$ & $(10^{42} {\rm erg/s})$ & $(10^{42} {\rm erg/s})$ & $(10^{42} {\rm erg/s})$ & $(10^{42} {\rm erg/s})$ \\
\hline
  51081&       7.1$\pm$  1.0&       9.3$\pm$  0.6&       4.0$\pm$  0.5&       1.3$\pm$  0.5&       2.7$\pm$  1.5\\
  51819&      29.8$\pm$  1.3&      39.9$\pm$  0.8&      17.4$\pm$  0.6&       9.2$\pm$  0.5&      16.4$\pm$  1.6\\
  52225&      20.9$\pm$  1.1&      26.5$\pm$  0.7&      10.5$\pm$  0.5&       4.9$\pm$  0.5&      11.0$\pm$  1.4\\
  52288&      17.1$\pm$  1.1&      21.9$\pm$  0.7&       9.5$\pm$  0.6&       5.5$\pm$  0.5&      11.5$\pm$  1.5\\
  52522&      13.7$\pm$  1.1&      14.0$\pm$  0.6&       7.1$\pm$  0.5&       4.0$\pm$  0.5&       7.4$\pm$  1.6\\
  52551&      13.6$\pm$  1.1&      11.4$\pm$  0.6&       5.2$\pm$  0.5&       1.6$\pm$  0.5&       6.1$\pm$  1.6\\
  52558&      12.9$\pm$  1.0&      13.5$\pm$  0.6&       5.3$\pm$  0.5&       3.1$\pm$  0.5&      -2.9$\pm$  2.5\\
  52576&      12.3$\pm$  1.0&      15.3$\pm$  0.6&       5.1$\pm$  0.5&       2.7$\pm$  0.5&       6.4$\pm$  1.6\\
  52577&      14.4$\pm$  1.0&      16.8$\pm$  0.6&       6.9$\pm$  0.5&       4.4$\pm$  0.5&       9.5$\pm$  1.5\\
  52585&      12.2$\pm$  1.1&      15.6$\pm$  0.6&       5.4$\pm$  0.5&       2.9$\pm$  0.6&       7.3$\pm$  1.8\\
  52909&      18.4$\pm$  1.1&      21.6$\pm$  0.6&       9.0$\pm$  0.5&       4.3$\pm$  0.5&      10.1$\pm$  1.5\\
  52910&      19.2$\pm$  1.1&      21.1$\pm$  0.7&       9.8$\pm$  0.5&       4.4$\pm$  0.5&       9.6$\pm$  1.5\\
  52935&      15.9$\pm$  1.1&      17.6$\pm$  0.7&       7.4$\pm$  0.6&       3.1$\pm$  0.6&       7.8$\pm$  1.7\\
  52963&      14.8$\pm$  1.0&      16.2$\pm$  0.6&       5.5$\pm$  0.5&       3.7$\pm$  0.5&       7.3$\pm$  1.4\\
  52971&      10.6$\pm$  1.1&      12.6$\pm$  0.6&       5.0$\pm$  0.5&       2.1$\pm$  0.5&       4.0$\pm$  1.5\\
  53351&       9.1$\pm$  1.0&       8.0$\pm$  0.6&       3.9$\pm$  0.5&       2.1$\pm$  0.6&       3.0$\pm$  1.7\\
  53634&       3.4$\pm$  3.0&       6.5$\pm$  1.7&       1.6$\pm$  1.1&       1.8$\pm$  0.7&       3.6$\pm$  1.8\\
  53655&       5.0$\pm$  1.1&       4.5$\pm$  0.7&       3.2$\pm$  0.6&       0.8$\pm$  0.5&       2.5$\pm$  1.7\\
  53665&       2.6$\pm$  2.1&       3.9$\pm$  1.0&       1.6$\pm$  0.7&       0.3$\pm$  0.6&       1.5$\pm$  1.8\\
  53669&       5.1$\pm$  1.1&       3.9$\pm$  0.6&       1.7$\pm$  0.5&       1.0$\pm$  0.5&       2.3$\pm$  1.6\\
  53671&       4.6$\pm$  1.4&       3.6$\pm$  0.7&       2.3$\pm$  0.7&       0.9$\pm$  0.6&       3.6$\pm$  1.7\\
  53676&       3.0$\pm$  1.2&       2.5$\pm$  0.6&       0.6$\pm$  0.6&       0.1$\pm$  0.5&       1.3$\pm$  1.5\\
  53679&       2.3$\pm$  1.4&       1.9$\pm$  0.7&      -0.8$\pm$  0.7&      -0.3$\pm$  0.6&      -0.0$\pm$  1.8\\
  53700&       4.9$\pm$  1.3&       4.7$\pm$  0.7&       2.0$\pm$  0.6&       1.2$\pm$  0.6&       6.4$\pm$  1.9\\
  53705&       3.5$\pm$  1.3&       2.8$\pm$  0.7&       2.0$\pm$  0.6&       0.9$\pm$  0.6&      -0.8$\pm$  1.7\\
  53975&       3.1$\pm$  1.2&       4.5$\pm$  0.6&       2.2$\pm$  0.5&       1.5$\pm$  0.5&       2.4$\pm$  1.7\\
  53994&       4.0$\pm$  1.2&       2.7$\pm$  0.6&       1.0$\pm$  0.5&       0.4$\pm$  0.5&      -3.0$\pm$  1.4\\
  54008&       2.8$\pm$  1.1&       3.0$\pm$  0.6&       1.8$\pm$  0.5&       1.4$\pm$  0.5&       1.4$\pm$  1.7\\
  54010&       2.7$\pm$  0.9&       3.1$\pm$  0.6&      -0.3$\pm$  1.0&       1.0$\pm$  0.5&       0.1$\pm$  1.4\\
  54012&       4.1$\pm$  1.8&       0.0$\pm$  1.1&      -1.6$\pm$  0.7&      -2.4$\pm$  0.6&       0.2$\pm$  1.7\\
  54020&       2.3$\pm$  1.7&       5.8$\pm$  1.0&       4.8$\pm$  0.7&       2.6$\pm$  0.7&       4.5$\pm$  2.0\\
  54024&       2.7$\pm$  1.5&       4.3$\pm$  0.7&       3.2$\pm$  0.6&       2.1$\pm$  0.6&       5.6$\pm$  1.8\\
  54030&       2.7$\pm$  0.9&       4.0$\pm$  0.6&       2.2$\pm$  0.6&       1.2$\pm$  0.5&       0.6$\pm$  1.6\\
  54037&       3.6$\pm$  1.0&       3.7$\pm$  0.6&       2.2$\pm$  0.5&       1.2$\pm$  0.5&       1.4$\pm$  1.4\\
  54040&       1.5$\pm$  1.5&       3.4$\pm$  1.0&       0.7$\pm$  0.8&       0.4$\pm$  0.6&       0.8$\pm$  1.7\\
  54049&       3.6$\pm$  2.2&       3.8$\pm$  1.1&       2.9$\pm$  0.7&       1.4$\pm$  0.6&       1.6$\pm$  1.9\\
  54053&       3.0$\pm$  1.1&       5.1$\pm$  0.7&       3.4$\pm$  0.6&       1.5$\pm$  0.6&       3.6$\pm$  1.7\\
  54056&       2.6$\pm$  1.3&       4.2$\pm$  0.6&       0.6$\pm$  0.6&      -0.8$\pm$  0.6&      -1.7$\pm$  1.7\\
  54059&       3.2$\pm$  1.3&       4.1$\pm$  0.7&       1.9$\pm$  0.6&       1.3$\pm$  0.6&       6.7$\pm$  1.9\\
  54062&       3.7$\pm$  1.2&       3.4$\pm$  0.6&       1.6$\pm$  0.6&       0.8$\pm$  0.5&      -0.6$\pm$  1.7\\
  54357&       3.5$\pm$  1.0&       3.6$\pm$  0.6&       1.8$\pm$  0.5&       1.2$\pm$  0.5&       1.9$\pm$  1.4\\
  54359&       2.3$\pm$  1.4&       3.7$\pm$  0.7&       1.6$\pm$  0.6&       0.7$\pm$  0.7&       0.2$\pm$  2.0\\
  54362&       3.3$\pm$  1.1&       3.3$\pm$  0.6&       1.7$\pm$  0.6&       1.7$\pm$  0.5&      -1.4$\pm$  1.5\\
  54376&       2.1$\pm$  1.2&       4.1$\pm$  0.7&       1.4$\pm$  0.6&       0.7$\pm$  0.5&       2.3$\pm$  1.8\\
  54382&       1.8$\pm$  1.1&       2.3$\pm$  0.6&       0.9$\pm$  0.6&       1.2$\pm$  0.6&       1.8$\pm$  1.8\\
  54385&       1.8$\pm$  1.1&       3.1$\pm$  0.6&       1.1$\pm$  0.5&       0.6$\pm$  0.5&       0.3$\pm$  1.4\\
  54411&       3.5$\pm$  1.0&       3.4$\pm$  0.6&       2.2$\pm$  0.5&       0.8$\pm$  0.5&       0.7$\pm$  1.4\\
  54413&       2.0$\pm$  1.4&       3.6$\pm$  0.7&       2.7$\pm$  0.6&       0.8$\pm$  0.6&       1.2$\pm$  1.9\\
  54415&       4.2$\pm$  1.1&       3.4$\pm$  0.9&       1.6$\pm$  0.5&       1.0$\pm$  0.5&       2.4$\pm$  1.6\\
  54422&       3.3$\pm$  1.3&       2.9$\pm$  0.6&       1.6$\pm$  0.6&       0.7$\pm$  0.6&       0.5$\pm$  1.6\\

\hline
\end{tabular}
\end{table*}

We then calculate the `baseline' non-varying host galaxy luminosity in each of the five 
SDSS filters by assuming that the weighted average of the light collected
in the last 10 epochs of Stripe-82 data (autumn 2007) is due to the large-scale galaxy stellar
emission, plus a small contribution from a nuclear power-law (which we constrained by analysing the 2010 BOSS spectrum, see section~\ref{sec:spectra} above). The SED of the baseline `pure' (i.e. with the nuclear power-law component removed) galaxy emission is shown in Fig.~\ref{fig:spectrum_lc_left} as empty stars.

We then derive the `nuclear' luminosity\footnote{All luminosities have been computed from the measured fluxes adopting a 
$\Lambda$CDM cosmology with $\Omega_{\rm m}=0.286$, $\Omega_{\Lambda}=0.714$ and $H_0=69.6$ km/s/Mpc.} 
in each SDSS filter by subtracting the baseline host flux from the observed 
flux at each Stripe-82 epoch (see Table~\ref{table:lc}, column (2) to (6), where we report the source nuclear luminosities based on SDSS {\tt modelMag}). 
Finally, we compute the total optical
luminosity of the nuclear flare 
$L_{\rm opt,nuc}=\int_u^z L_{\lambda} d\lambda \simeq \sum_{i} L_{\lambda,i} \Delta \lambda_i$, 
where the integral extends over the entire SDSS bandpass, and the 
summation is over the {\it u'g'r'i'z'} SDSS filters, each characterized by a bandwidth $\Delta \lambda_i$ \citep{fukugita96}.

\begin{figure*}
\includegraphics[width=0.75\textwidth,clip]{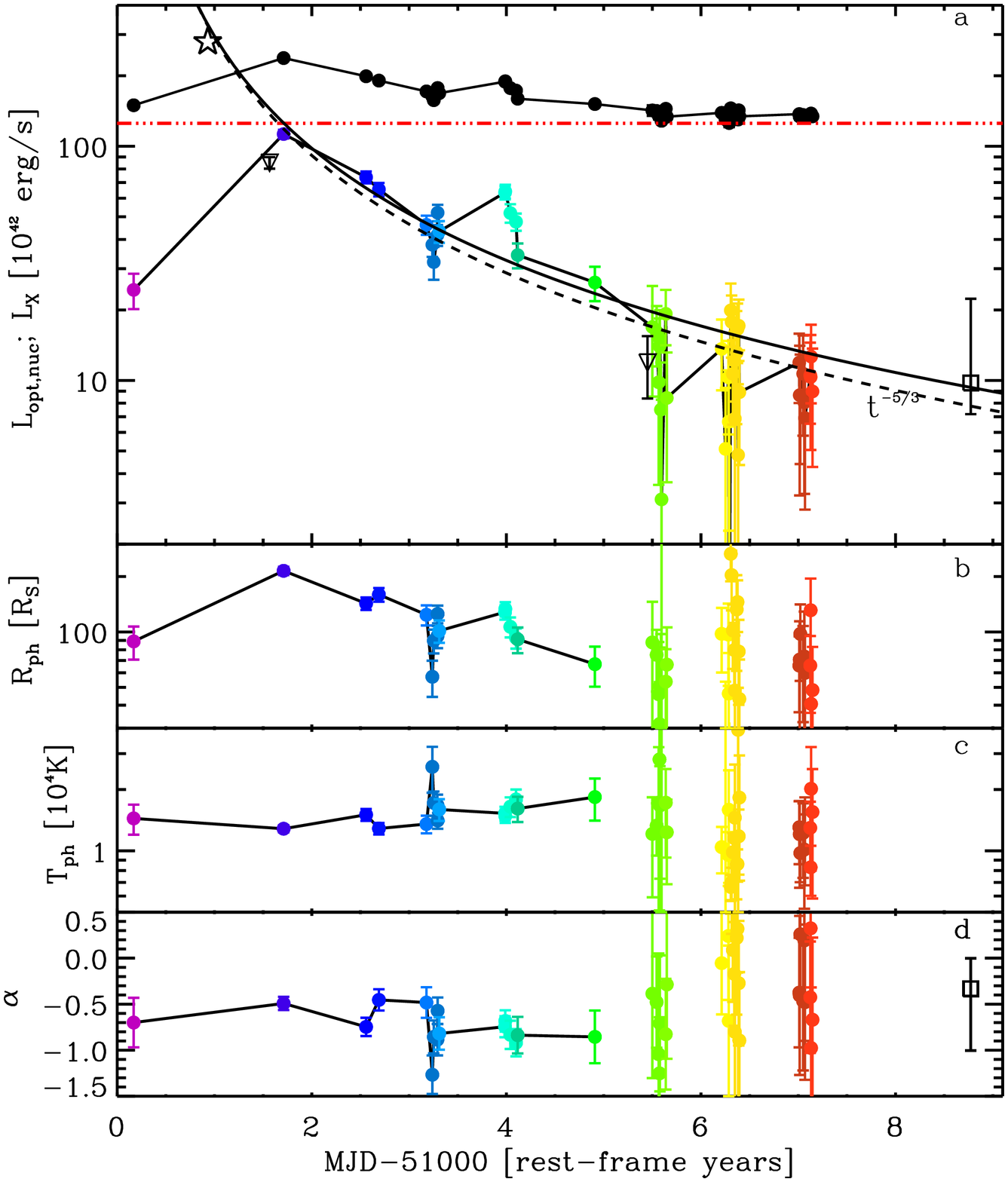}
\caption{{\it Panel a}: the time evolution of the nuclear luminosity is shown with the coloured circles (where each colour identifies one epoch of the SDSS
Stripe 82 observations). The black circles are the total optical luminosity of SDSS~J0159+0033, shown here for reference, together with the 
estimated (constant) luminosity of the host galaxy (red horizontal dot-dashed line). The empty triangles show the X-ray (2--10 keV) luminosity 
measured by {\it XMM-Newton} (first epoch) and {\it Chandra} (second epoch). The black solid line is the best fit power-law evolution of the flare decay, including all data points (with slope of $-1.59\pm0.05$). The dashed line, which is not a fit to the data, shows the $t^{-5/3}$ 
long-term decline expected in most TDE models. The empty star shows the approximate location of the peak in our ``fiducial'' TDE model 
(see text for details), while the empty square at late times is the corresponding optical luminosity of the power-law component fitted to the BOSS 2010 spectrum,
as described in section~\ref{sec:spectra}, and plotted in Fig.~\ref{fig:spectrum_lc_right} with a purple dashed line. 
{\it Panel b and c} show the evolution of the best fit photospheric radius (in units of the Schwarzschild radius 
for a $10^{8} M_{\odot}$ black hole) and temperature (in units of $10^{4}$ K), obtained by fitting the optical photometric SED with a simple black-body spectrum (filled circles). {\it Panel d} shows the 
best-fit power-law slope $\alpha$, where $L_{\lambda} \propto \lambda^{-2+\alpha}$. In all panels, the black solid line connecting the points is drawn to guide the eye.}
\label{fig:nuc_lc}
\end{figure*}

The top panel of Fig.~\ref{fig:nuc_lc} shows the time evolution of the optical nuclear luminosity $L_{\rm opt,nuc}$, 
where each epoch is represented by a data point of a different colour. A simple power-law fit to the time evolution of the flare in its decay phase 
gives a slope of $n_{\infty}=-1.59\pm0.05$, only slightly shallower than 
the predictions of simple tidal disruption flare models \citep[a power-law evolution in with slope $-5/3$][]{rees88,lodato09,guillochon13}.

\begin{figure}
\includegraphics[width=0.5\textwidth,clip]{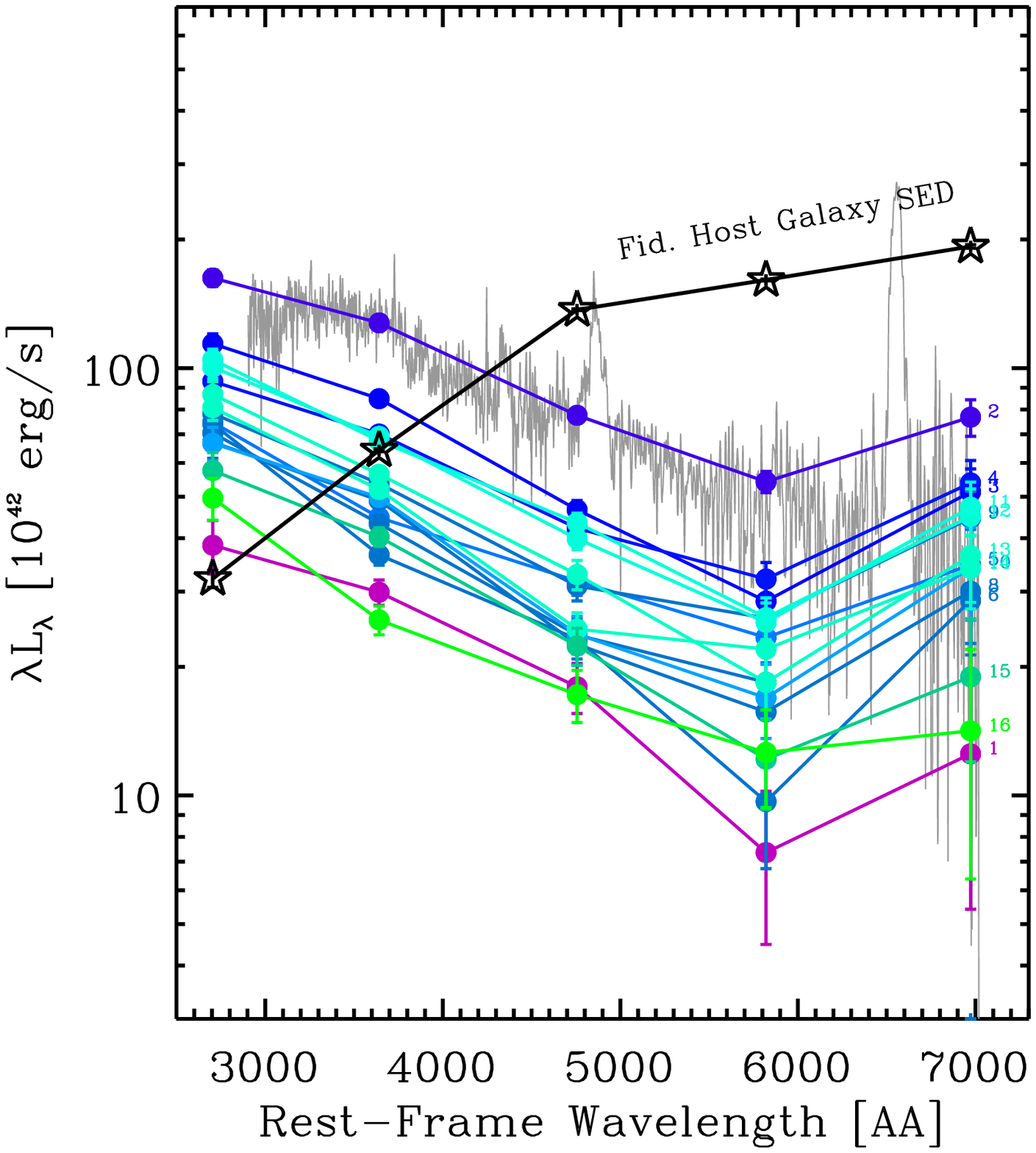}
\caption{Time evolution of the nuclear optical SED during the flare of SDSS~J0159+0033. For each of the 
first 16 epochs of the cleaned Stripe-82 light-curve, spanning the first 5 rest-frame years, shown in Fig.~\ref{fig:nuc_lc}, we plot the monochromatic luminosity, 
$\lambda L_{\lambda}$, in units of $10^{42}$\,erg\,s$^{-1}$, in the {\it u'g'r'i'z'} bands as filled coloured circles. 
The colour-code is identical to that of Fig.~\ref{fig:nuc_lc}: each colour corresponds to a different Stripe-82 epoch, as marked on the right of each SED. 
As in Fig.~\ref{fig:spectrum_lc_left}, the empty black stars indicate the baseline (constant) host galaxy SED, derived by taking the weighted average 
of the photometric data-points of the last season of Stripe-82 observations (last ten epochs, Autumn 2007) 
and subtracting the small power-law contribution that was estimated from the 2010 spectrum.
The light grey line shows the 
difference spectrum between the peak flare (2010) and the host, with the corresponding difference-photometry data-points marked as blue solid circles 
(see section~\ref{sec:bol_lc} for more details).}
\label{fig:spectrum_lc_right}
\end{figure}

Fig.~\ref{fig:spectrum_lc_right} shows the evolution of the optical SED of the flare; 
for clarity, only the first 16 epochs of Stripe-82 photometry are shown, with the same colour code of Fig.~\ref{fig:nuc_lc}.
As a reference, the plot shows also the constant baseline host galaxy spectral energy distribution in the five optical bands 
(black empty stars, as in Fig.~\ref{fig:spectrum_lc_left}).
The emission in the {\it u'g'r'i'} filters is consistent with a spectrum that rises steeply with frequency. On the other hand, the {\it z'} band
magnitudes are affected by the strong $H_{\alpha}$ emission, which, at the peak, contributes to about 25\% of the observed flux in that band, and this
contribution increases as the luminosity of the power-law continuum declines.
To better characterize the spectral evolution of the flare, we first fitted 
the {\it u'g'r'i'} data points with a simple power-law $L_{\nu} \propto \nu^{-\alpha}$ 
(or, equivalently, $L_{\lambda} \propto \lambda^{-2+\alpha}$), and we plot the evolution of the spectral index $\alpha$ in panel (d) 
of Fig.~\ref{fig:nuc_lc}. The slopes that we obtain are slightly steeper than 
the expected $\nu^{1/3}$ law of a geometrically thin, optically thick (un-truncated) accretion disc \citep{ss73}, but are consistent
with previous optical observations of well-sampled TDEs \citep{gezari12,holoien14}.
The overall `colour' evolution of the flare, as diagnosed by the Stripe-82 {\it u'g'r'i'} light-curves, is mild, and reminiscent of the almost 
constant-temperature evolution seen in most well-sampled TDE light-curves \citep{gezari12,holoien14}. In fact, this behaviour is a challenge
to the simplest model of a `bare' viscous accretion disc evolution for the decay phase of tidal disruption events, and \citet{guillochon14} 
have argued that an extended, large scale reprocessing layer should be present to produce the observed (almost) achromatic evolution of TDE flares.
Such a reprocessor should be expected on the basis of the results of hydrodynamical simulations, that show indeed large amounts of debris
present at distances ranging from the outer disc of the bound material that slowly returns to pericenter, $r_0 \approx 2(GM_{\rm BH}/\pi^2)^{1/3} t^{2/3}$, to the size of the expanding shell of unbound material expelled from the disrupted portions of the star, which moves away from the black hole (see section~\ref{sec:tde} below).
As a simple phenomenological test of the above scenario, 
we have also fitted the same spectral evolution data shown in fig.~\ref{fig:spectrum_lc_right} with a single-temperature black-body spectrum, assuming, for simplicity, an 
emitting area $A_{\rm ph}=4\pi R_{\rm ph}$. Panel (c) of Fig.~\ref{fig:nuc_lc} shows the evolution of the best-fit photospheric temperature
(in units of $10^4$ K), while panel (b) shows the time evolution of the photospheric radius of the emitting surface, $R_{\rm ph}$ (in units of 
the Schwarzschild radius for a $10^8 M_{\odot}$ black hole). Indeed, photospheric temperature changes throughout the flare appear to be mild, if not negligible: the flare displays both temporal and spectral evolution very similar to known UV/optical tidal
disruption flares. We discuss in the following section the constraints on the TDE model parameters we can obtain by modelling these data.

\section{The flare as a tidal disruption event}
\label{sec:tde}
A star of mass $M_{*}$ and radius $R_{*}$ that happens to reach 
a distance from a supermassive black hole of mass $M_{\rm BH}$ of the order of the
star's tidal radius $R_{\rm T}\simeq R_{*} (M_{\rm BH}/M_{*})^{1/3}$, will not survive the encounter unscathed.
Depending on the star's inner structure and on the penetration factor $\beta \equiv R_{\rm T}/R_{\rm p}$ (where $R_{\rm p}$ is the pericenter radius), 
various degrees
of disruption will be unavoidable, with deep encounters ($\beta>1$) causing the star to be completely torn apart \citep{ayal00,guillochon13}.
We can rewrite the tidal radius in units of the black hole's Schwarzschild radius ($R_{\rm S} = 2GM_{\rm BH}/c^2$):
\begin{equation}
R_{\rm T}/R_{\rm S}\simeq 5.06 \left(\frac{M_{*}}{M_{\odot}}\right)^{-1/3}\left(\frac{M_{\rm BH}}{10^7 M_{\odot}}\right)^{-2/3}\left(\frac{R_{*}}{R_{\odot}}\right) 
\label{eq:rt_rs}
\end{equation}
If the black hole is massive enough, or the star compact enough, then we expect tidal disruption will only take place inside the event horizon (located at $R_{\rm S}$ for non-spinning black holes, and at $R_{\rm S}/2$ for maximally spinning Kerr black holes), and will be invisible to the outside observers. More
accurate General Relativistic calculation \citep[see e.g.][]{kesden12b} will then be needed to predict the incidence of TDEs from spinning black holes.

In tidal disruption events, whatever fraction of the star's mass is shed by black hole tidal forces, 
about half of it will remain bound to the central black hole.
After an initial short `fall-back' time, given by the time it takes the most bound material to return to pericenter assuming the star is initially on a 
parabolic orbit \citep{rees88,macleod12}:
\begin{equation}
t_{\rm fb}\simeq 0.37 \beta^{-3}  \left(\frac{M_{*}}{M_{\odot}}\right)^{-1}\left(\frac{M_{\rm BH}}{10^7 M_{\odot}}\right)^{1/2}\left(\frac{R_{*}}{R_{\odot}}\right)^{3/2} {\rm yrs}, \,
\label{eq:tfb}
\end{equation}
the stellar debris will pile up near pericenter. 
A number of possible mechanisms (general relativistic precession, hydrodynamical dissipation, compressive magneto-rotational instability, 
see \citealt{guillochon14}) could then lead to rapid dissipation of the material's internal energy, circularizing the orbit of the debris. 
Viscous dissipation will then lead to accretion onto the black hole. Provided that both dissipation and viscous transport of angular 
momentum can act efficiently, the rate of mass
accretion onto the black hole is then fixed by the rate of mass return to pericenter. This, in turn, can be computed using Kepler's third law \citep{rees88}, 
and depends critically on the specific binding energy of the stellar material at the time of disruption (d$M$/d$E$). 
For flat distributions (d$M$/d$E=$ constant), this leads to the well-known $t^{-5/3}$ evolution of the accretion rate at pericenter. 
Detailed numerical simulations of TDEs have mostly confirmed that this
is indeed to be expected, at least in a `bolometric' sense, and at times longer than the peak accretion time \citep[see e.g.][]{ramirez09,lodato09,guillochon13,guillochon14}.
In fact, \citet{guillochon13} have also performed a detailed parameter study of the expected light-curve of different TDEs as a function of 
black hole mass, stellar mass, penetration factor and the star's adiabatic index. In particular, they have shown how the peak accretion time and rate, as well as the asymptotic slope, $n_{\infty}$, of the accretion rate light-curve can all be used to put constraints on the parameters of the encounter, including the masses of the star and the
hole, and $\beta$ \citep{guillochon14}. In the following, we will use the scaling derived from these parametric study, but we 
stress that the issues of debris circularization, and the efficiency with which the rate of matter returning at pericenter 
material can be converted into an accretion rate onto the black hole are currently open ones, and subject to a intense study 
\citep{hayasaki15,shiokawa15,guillochon15}.

\subsection{Constraining the parameters of TDE models}
\label{sec:model}
Based on the outcome of hydrodynamical simulation of (Newtonian) TDEs, \citet{guillochon13} have presented scaling relations for the most 
important light-curve parameters, as a function of black hole mass, stellar mass and radius, polytropic index of the stellar structure and penetration factor. In particular, they showed that the peak accretion rate scales as:
\begin{equation}
\label{eq:mdot_peak}
\dot M_{\rm peak} = A_{\gamma} \left(\frac{M_{\rm BH}}{10^6 M_{\odot}}\right)^{-1/2} \left(\frac{M_*}{M_{\odot}}\right)^{2} \left(\frac{R_{*}}{R_{\odot}}\right)^{-3/2} M_{\odot}{\rm /yr}\;
\end{equation}
while the peak time of the accretion rate follows:
\begin{equation}
\label{eq:t_peak}
t_{\rm peak} = B_{\gamma} \left(\frac{M_{\rm BH}}{10^6 M_{\odot}}\right)^{1/2} \left(\frac{M_*}{M_{\odot}}\right)^{-1} \left(\frac{R_{*}}{R_{\odot}}\right)^{3/2} {\rm yrs}.
\end{equation}

In the above expressions, $A_{\gamma}$ and $B_{\gamma}$ represent rational functions of the penetration factor $\beta$, evaluated from the numerical simulations for different values of the polytropic index $\gamma$ (assumed to be equal to $4/3$ and $5/3$ for high- and low-mass stars, respectively). Their
form can be found in Eqs. A5-A8 of \citet{guillochon13,guillochon15e}. To further simplify the analysis, we assume a fixed mass-radius relation for main-sequence 
stars, as given by \citet{tout96}; we fix the boundary between low- and high-mass stars at 0.6$M_{\odot}$, and the black hole mass to 
$M_{\rm BH}=10^8 M_{\odot}$, so that the overall family of TDE light-curve only depends on the penetration factor $\beta$ and the 
star's mass $M_{*}$.

The allowed region in this two-dimensional parameter space is determined by imposing two conditions\footnote{Because the SDSS Stripe-82 
light-curve did not well sample the rising phase of the outburst, and we do not know the exact time of the 
star's passage at pericenter, we cannot impose a third constraint on the peak time. As a reference, the 
time elapsed between the first and the second SDSS photometric data points is approximately 1.176 rest-frame years.}: the first is 
that the accretion rate at the peak is sufficiently high to explain 
the observed peak luminosity of the nucleus, $L_{\rm opt,peak}\simeq 1.1 \times 10^{44}$ erg/s, 
(for a given radiative efficiency, $\epsilon$ and optical-to-bolometric correction $\kappa_{\rm opt}$):
\begin{equation}
\label{eq:mpeak_limit}
\dot M_{\rm peak} \ge 9.8 \times 10^{-2} \left(\frac{\kappa_{\rm opt}}{5}\right)\left(\frac{\epsilon}{0.1}\right)^{-1} M_{\odot}/{\rm yr}.
\end{equation}
The second is that the pericenter passage must occur outside the event horizon. For simplicity, we assume a non-spinning black hole, 
thus $R_{\rm p}/R_{\rm S}=\beta^{-1}R_{\rm T}/R_{\rm S}\ge 1$ (see Eq.~\ref{eq:rt_rs} above). For Kerr black holes, closer encounters are 
possible, and for maximally spinning black holes, the condition would read $R_{\rm p}>R_{\rm g}=R_{\rm S}/2$. 

\begin{figure*}
\includegraphics[width=0.9\textwidth,clip]{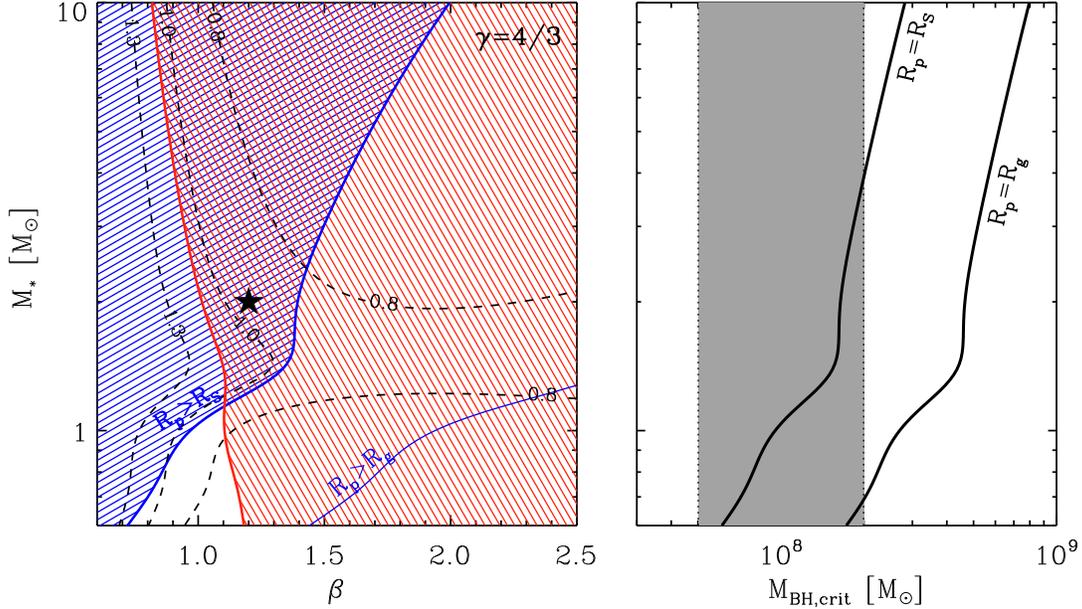}
\caption{{\it Left}: The allowed range of the TDE model parameters $\beta\equiv R_{\rm T}/R_{\rm p}$ (penetration factor) and $M_{*}$ (star's mass) is shown as a 
doubly shaded (blue and red) wedge. It is derived by the combination of two independent constraints: (i) pericenter passage must occur outside the SMBH event 
horizon ($R_{\rm p}>R_{\rm S}$, thick blue line, for the two extreme cases of non-spinning black hole and thin blue line for the case of maximally spinning hole, 
$R_{\rm p}>R_{\rm g}$); and (ii) the peak luminosity of the flare has to be at least as high as the observed one
 (assuming a optical-to-bolometric correction of 5 and a radiative efficiency of 10\%, solid red line). 
The black dashed lines show the contours of equal peak time, in rest-frame years. 
The black star mark a `fiducial' model, discussed in the text, for illustration purposes. {\it Right:} The relation between the critical black hole mass $M_{\rm BH,crit}$ (x-axis), above 
which stars are swallowed whole and no tidal disruption flare can be observed and the star's mass (y-axis) is shown as a black solid line for the case of
non-spinning black holes (leftmost line) and maximally spinning black holes (rightmost line). The vertical dashed area mark the range of possible SMBH
masses estimated for SDSS~J0159+0033 on the basis of the observed broad emission lines.}
\label{fig:good}
\end{figure*}

The left panel of Fig.~\ref{fig:good} shows the allowed range of the two-dimensional parameter space defined by 
the penetration factor $\beta$ and the star's mass (in solar units), given the above constraints.
We obtain a minimum mass of about $M_{*,{\rm min}}=1.2 M_{\odot}$ and a quite narrow range of penetration factors, which widens up as more massive stars
are considered. Just as a reference point for discussion, we mark with a black star in Fig.~\ref{fig:good} a `fiducial' parameter combination 
allowed by the data: a 2 solar mass star with pericenter passage just 1.2 times smaller than its tidal radius. Such an event would produce a flare
of optical luminosity $L_{\rm opt,peak}\simeq 2.8 \times 10^{44} (\epsilon/0.1) (\kappa_{\rm opt}/5)^{-1}$, almost three times higher than the maximum
caught by SDSS in 2000. The peak time would be $\approx$0.96 years after disruption, and, from Eqs. (A3-A4) of \citet{guillochon13}, 
would evolve to 
the asymptotic decay power-law slope $n_{\infty}\approx 1.64$, consistent, within the uncertainties, with the observed value. 
Based on the same fiducial model, the amount of mass
lost by the star would be $\Delta M \approx 0.36 M_{\odot}$ ($\approx 18$\% of the star's mass).
Indeed, as a final consistency check, we derived the (bolometric) fluence of the flare, measured with only the available data points: 
${\cal F}_{\rm bol}\approx 4\times 10^{52} (\kappa_{\rm opt}/5)$ ergs. 
Assuming a 10\% radiative efficiency, this corresponds to about 0.2 $M_{\odot}$ accreted onto the central 
super-massive black hole, consistent with the expectations of the fiducial TDE model described above. 

The right panel of Fig.~\ref{fig:good} shows the relation between the critical black hole mass $M_{\rm BH,crit}$, above 
which stars are swallowed whole and no tidal disruption flare can be observed and the star's mass, for the case of
non-spinning black holes and maximally spinning black holes.

The above estimates serve to emphasize, first of all, that a TDE explanation for the observed flare 
is energetically viable, and, secondly, that well sampled light-curves of tidal disruptions flares by black holes of known mass could provide 
tight constraints on accretion parameters that are typically elusive in steady accreting systems, such as the radiative efficiency 
(and thus the black hole spin), and the accretion flow bolometric corrections \citep{guillochon14}.

To conclude this section, we note that a prominent re-brightening flare was observed about 4 rest-frame years after the peak.
This sub-flare is characterized by a relatively constant continuum slope and photospheric temperature, and has a fluence of about 
$5.5 \times 10^{51} (\kappa_{\rm opt}/5)$ ergs, corresponding to 14\% of the fluence of the whole flare. 
Among well-monitored TDE so far, only
the {\it Swift}-selected `relativistic' events \citep{burrows11,bloom11,cenko12} did show clear structures in their 
declining light-curves, but it is possible that 
this is simply a selection effect due to lack of long-term dense monitoring of more recently discovered optical TDEs. 
Lacking a large sample of well sampled optical light-curves of TDEs,  
it is beyond the scope of this paper to speculate about the nature and causes of this apparent secondary flare event, but
one should consider testing hydrodynamical models of tidal disruptions, including the expectations from binary events \citep{mandel15}, 
to assess whether such events may pose a real challenge to the TDE 
interpretation of this flare, and point towards an alternative explanation in terms of AGN accretion physics, as we discuss below.

\section{Discussion}
\label{sec:discussion}

A previous search for TDEs in the Stripe-82 data \citep{vanvelzen11} did not select SDSS~J0159+0033 as a good tidal disruption candidate. 
We believe there are two concurrent 
reasons for this: first of all, the flare we observe is rather slow (taking more than 5 years to return to quiescence), 
whilst \citet{vanvelzen11} selected
preferentially flares which lasted only one season of Stripe-82 observations. More critically, \citet{vanvelzen11} excluded from their final list of TDE candidates all those with
known AGN spectroscopic classification. It is just due to a fortuitous coincidence that the first 2000 SDSS spectrum, showing 
the prominent broad emission lines that led to type-1 AGN classification, was taken very close in time to the flare 
peak, and we suspect this crucial detail had been overlooked. 

Of course, the main question we would like to answer is the following: is the observed flare from SDSS~J0159+0033 the result of the tidal disruption of a star, or just
a particularly strong and rapid AGN outburst? To a certain extent, this is an ill-posed question, as TDE {\it are} in fact just a specific class of AGN outbursts.
The analysis presented in the previous section, however, suggests that the observed light-curve evolution, and, in particular, the overall energetics of the event
are consistent with the amount of accreted mass being a substantial fraction of a massive main-sequence star. 
In general terms, well-sampled light-curves are probably the most powerful tool in order to distinguish TDE-induced
flares from stochastic AGN variability, as demonstrated by the recent example of the nearby source IC 3599 \citep{campana15}.

Irrespective of the source of the accreted material (star or gas cloud, for example), it is clear that it must have been deposited very close to the central 
black hole. If the accretion energy we see was generated in a standard, viscous, \citet{ss73} disc, the viscous (or infall) time, can be written as:
\begin{equation}
t_{\rm visc}\simeq 1.7 \times 10^{-4} m_8 \alpha_{0.1} \left(\frac{H}{R}\right)^{-2} r^{3/2},
\end{equation}
where $m_8$ is the BH mass in units of $10^{8} M_{\odot}$, $\alpha_{0.1}$ is the viscosity parameter in units of $0.1$ and $r$ is the radius of the disc 
in units of $R_{\rm g}$. The disc scale-height in the radiation-pressure dominated inner part of the disc, which is the region relevant for the luminous accretion episode discussed here, can be expressed as $H/R \simeq 20 \lambda_{\rm Edd} J(r)/r$, where $\lambda_{\rm Edd}\equiv L_{\rm bol}/L_{\rm Edd}$ is the Eddington ratio 
($\lambda_{\rm Edd}\approx 0.04 (\kappa_{\rm opt}/5)$ at the observed peak, and a factor of three higher at the fiducial peak) 
and $J(r)=1-\sqrt{r_{\rm in}/r}$ is the Newtonian correction factor from the no-torque inner boundary condition.
Thus, viscous times shorter than one year require disc sizes smaller than 10 gravitational radii (i.e. smaller than a few times the pericenter distance, with
$R_{\rm p}\approx 3.3 R_{\rm g}$ in our fiducial model), with the exact details dependent on the disc inner boundary condition (and BH spin).
Progressive draining of a thin disc substantially larger than this is plainly incompatible with the observations.
However, thermal instabilities can also operate in the inner parts of the disc, which will be faster by a factor of the order $(R/H)^2$. 
In fact, there are a number of potential mechanisms that theories of accretion discs have put forward to explain very rapid (faster than viscous) variability in the inner region of an accretion discs, such as thermal instabilities, large scale waves in the inner accretion disk, hydromagnetic winds, reprocessing of UV or X-ray light \citep[][and references therein]{krolik05,lamassa15}.

The X-ray spectrum observed by {\it XMM-Newton} close to the peak of the lightcurve is consistent with a power-law of index 2.1 \citep{lamassa15}, similar to what typically observed
in luminous AGN. Simple thermal models for the emission from a TDE do not predict such a 'corona-like' spectrum, very much as the standard Shakura-Sunyaev theory of
geometrically thin and optically thick accretion discs does not predict the ubiquitous Comptonising medium inferred from X-ray spectra of AGN. Indeed, most
TDE candidates detected by {\it ROSAT}, {\it XMM-Newton} and/or {\it GALEX} showed mostly very soft spectra, 
well-fit by a blackbody model and/or much steeper power-law  
\citep{brandt95,bade96,komossa99,greiner00,gezari08,esquej08}. But harder X-ray spectra have been observed recently from candidate TDEs, too 
\citep{cenko12b,nikolajuk13,saxton14}. In general, too little is currently known about the spectral formation mechanisms in TDE, across the
entire electromagnetic spectrum, to be able to reject the TDE interpretation on the basis of the X-ray spectral shape only.
Moreover, we note that, in general, optically selected TDE tend to have low photospheric temperatures and be X-ray faint, 
while X-ray selected TDE tended to be hot (when thermal) and optically faint
\citep[see e.g. Fig. 4 in][]{gezari12b}, clearly suggesting we are barely scratching the surface in our understanding of TDE selection effects.

There are, moreover, other pieces of evidence that are not straightforward to interpret within the simplest TDE scenario we have described so far. In particular,
the properties of the emission lines observed in the spectra are not easily explained by a TDE model, as we discuss in the following subsections.

\subsection{The broad line region}
\label{sec:blr}
Only since the discovery of the first optically selected TDEs \citep{vanvelzen11,gezari12}, and the consequent availability
of early spectroscopic follow-up observations, it has become evident that the emission from tidal disruptions of stars is
accompanied by the presence of broad emission lines in their optical spectra. 
The broad emission line phenomenology is diverse, with some objects
showing only high-ionization lines (typically HeII $\lambda$4686), others only Balmer lines, and 
others still both He and H lines \citep{arcavi14}. Measured line widths vary between a few thousands to about ten thousand \kms.
If these broad emission lines are produced by stellar debris illuminated by the tidal disruption flare, we can immediately rule out
them being located at a distance from the hole of the order of the circularization radius (typically assumed to be equal to $2R_{\rm p}$),
as this would imply (Keplerian) velocity widths of a few times $10^4$ \kms. In fact, two possible alternative sites of broad emission line
production have been invoked for TDE. \citet{guillochon14} argues that the bound material extending out to $r_0$
(see section~\ref{sec:model} above) will constitute an elliptical accretion disc, growing inside-out,  eventually
reaching densities and ionization states leading to the generation of broad, permitted atomic emission lines, very much like those 
generated by long-lived AGN. According to this scenario, high-ionization lines should appear first, followed by lower ionization lines later on
during the outburst. Alternatively, \citet{strubbe09} considered the outbound stellar material, flying away from the disruption site
and extending out to $R_{\rm max}= R_{\rm T} \beta^{1/2} (t/t_{\rm dyn,*}) \approx 2.2 \times 10^{17} (R_{\rm p}/R_{\rm S})^{-1/2} [(t-t_{\rm p})/yr]$ cm, as the site of line emission processes. They, however, predicted both emission and absorption lines, and 
significant bulk blue-shift of the lines, contrary to what is typically observed. Moreover, as discussed in detail in \citet{guillochon14}, Newtonian 
hydrodynamic simulations clearly
show that the unbound material will be confined to a thin filament, covering a small solid angle as seen by the central black hole \citep{kochanek94}, 
hardly large enough to 
produce emission lines with the observed equivalent width.

The broad Balmer emission lines (H$\alpha$, H$\beta$ and H$\gamma$) observed in the 2000 SDSS spectrum of SDSS~J0159+0033 do appear at first sight 
like ordinary QSO broad lines: they are not strongly asymmetric, have no absorption through or P-Cygni profile, and have measured FWHM
ranging between 3.4 and 4.5 $\times 10^3$ \kms\ (see Table~\ref{tab:lines}). 
If they are produced by a mechanism analogous to that which generates the broad line region (BLR) in most AGN (an assumption we have implicitly used by adopting the ``single epoch'' virial black hole mass estimates for this object),
then they should follow the BLR luminosity-size relation ($R_{\rm BLR} \propto L^{\delta}$, with $\delta=0.4-0.6$), which would put them at a distance of about 30 
light days \citep{bentz06}, i.e. $\approx 2.6 \times 10^3 R_{\rm S}$ for a $10^8 M_{\odot}$ 
central black hole\footnote{Similar values for the BLR size would be obtained by adopting the empirical scaling between X-ray 2-10 keV luminosity 
and BLR size, \citet{greene10}.}. This is inconsistent with the \citet{guillochon14} `inner accretion disc' 
interpretation, as $r_0$ 
is only about half this value at the time of the 2000 SDSS spectroscopic observation, and only reaches these distances about three rest-frame 
years after the observed peak.

However, the most serious problem for the interpretation of the broad emission lines as originating in the stellar debris, are the implied gas mass by the standard
ionization/recombination model for BLR. Comparing the measured equivalent widths of \Ha\ and \Hb\ in the 2000 SDSS spectrum with those predicted
by \citet{korista04} as a function of the ionizing flux and BLR particle density, we derive a particle density of about log $n\approx 10.5$ 
(assuming an ionizing luminosity of $10^{44}$\,erg\,s$^{-1}$, and a distance of the BLR of 30 light days). 
The amount of material needed, assuming spherical geometry is a few hundred solar 
masses (see \citealt{baldwin03}), obviously much more than could have provided by the disrupted star. 

For the particle densities inferred for the BLR emitter, recombination is very fast, so the lines respond effectively instantaneously 
to flux variations from the central source. Thus, the rapid luminous flare we observed between 2000 and 2005, must have illuminated a pre-existing structure, whose
spatial distribution and kinematics closely resemble those of BLR of actively growing SMBH. Further evidence for this can be derived by the fact,
noticed by \citet{lamassa15}, that the measured H$\alpha$ FWHM between 2000 and 2010 scales as the continuum luminosity to the $-1/4$ power, 
as expected if gas in the BLR moves with Keplerian velocity.
 
The likelihood of the observed configuration, with the broad emission lines appearing transiently in response to a rapid nuclear TDE, depends on 
the lifetime of the broad-line region structure in objects where the 
central black has ceased accreting at substantial rates. This must depend, in turn, 
on the dynamical status of the BLR itself (is it outflowing, inflowing, circulating?), and,
critically, on the time elapsed since the last major accretion episode onto the SMBH. A detailed analysis of these issues is beyond the scope of this paper. In any case, the conclusion that such a configuration is possible 
was reached recently also by \citet{denney14}, who studied the
peculiar, decades-long flare of the central black hole in Mrk\,590, which showed dramatic brightening and subsequent disappearance of strong broad emission lines 
over little more than 20 years. 

\subsection{The narrow line region}
\label{sec:nlr}
It is not just the properties of the observed broad emission lines that suggest the flare we observed in SDSS~J0159+0033 occurred in a formerly active galaxy.
The system in question is massive and strongly star-forming, as indicated both by the SED fitting and by the strong 
\OII\ emission line detected (see section~\ref{sec:spectra} and Table~\ref{tab:lines} above). In the nearby universe, these are the systems most likely to host an AGN 
\citep{kauffmann03}. The narrow emission line diagnostics (see Fig.~\ref{fig:bpt}) 
reveal that the emitting material was probably ionized by a source harder than
that associated with pure star-formation. AGN activity would of course provide a sufficiently hard ionizing source, 
but the exact amount of any putative AGN contribution to the observed narrow 
line emission is harder to disentangle.
SDSS J0159+0033 appears to be located close to the boundary of the so-called `transition' region of parameter space, between the 
objects whose line ratios imply stellar sources
dominate the ionization, and those for which AGN emission is needed to produce the observed line ratios. \citet{kauffmann03}
 estimate that
the AGN fractional contribution for objects in the transition region ranges between 30 and 90 percent. In the most extreme case, then 
we can estimate the maximum luminosity that any putative AGN must have had in the past to produce 90 percent of the observed \OIIIb\ flux 
(corresponding to $L_{\rm [OIII]} \simeq 3-5 \times 10^{41}$ erg\,s$^{-1}$).

There is a large body of literature on the correlation between \OIIIb\ and hard X-ray luminosity in AGN, with somewhat contrasting results on the
exact numerical value of the correlation index and on the mean luminosity ratio. In general the ratio of observed 
(not extinction corrected) \OIIIb\ to X-ray (2--10 keV) luminosity ranges between 0.02 and 0.08, depending on the sample selection and source classification 
\citep{mulchaey94,heckman05,trouille10,degasperin11}. Thus, we can conservatively say that the central AGN must have had an X-ray luminosity  
 $L_{\rm 2-10 keV} \le 2 \times 10^{43}$ erg\,s$^{-1}$ (i.e. equal or smaller than the X-ray luminosity measured by {\it Chandra} in 2005, 
$L_{\rm 2-10 keV}=1.2 \times 10^{43}$ erg\,s$^{-1}$)
about one narrow-line region light-crossing time ago. Using the empirical relation between NLR size
and \OIIIb\ luminosity \citep{greene11,hainline13}, we derive a NLR size of about 4 kpc, corresponding to a look-back time of 
about 13 thousand years. As discussed in section~\ref{sec:lc}, the historical optical data pre-flare (i.e. before 1998) are consistent with
no strong AGN emission being present above the host galaxy stellar emission, but the uncertainty on the measurements is such that a low-level 
nuclear activity cannot be ruled out either, so that it is not possible to firmly constrain the possible AGN activity level in the galaxy prior to 
flare discussed here.

 Powerful TDE, just like longer episodes of nuclear activity powered by gas accretion,
should imprint in the host ISM an ionization `echo' \citep[see e.g.][]{yang13} detectable as high-ionization narrow emission lines, variable
on timescales longer than the light-crossing time at the location of their production.
Variability of strong coronal emission lines, possibly associated to the fading echo of a nuclear tidal disruption event has been
reported by \citet{komossa09,wang12,yang13}.
Systematic spatially-resolved spectroscopic surveys with integral field unit of nearby galaxies 
\citep[such as SAMI or MaNGA,][]{croom12,bundy15} could be able to detect signatures of such echos over the entire body of nearby galaxies,
thus probing up to timescales of the order $t_{\rm echo}\approx R_{\rm eff}/c \simeq 3\times 10^4 (R_{\rm eff}/10 {\rm kpc})$ yrs. Given
current estimates of the intrinsic rate of tidal disruption events per galaxy $\Gamma_{\rm TDE}\equiv 1/t_{\rm rec} \approx 10^{-5}$ yr$^{-1}$
(where we have introduced the recurrence time $t_{\rm rec}$; see sections~\ref{sec:intro} above and~\ref{sec:rates} below, 
and references therein), it is possible that a large fraction of nearby galaxies, of the order
of $t_{\rm echo}/t_{\rm rec}\approx 30$\%, could be found displaying off-center high-ionization emission line regions, 
thus probing the triggering properties of nuclear black holes, and of tidal disruption events (which have the shortest recurrence times) 
in particular.

\begin{figure*}
\includegraphics[width=0.99\textwidth,clip]{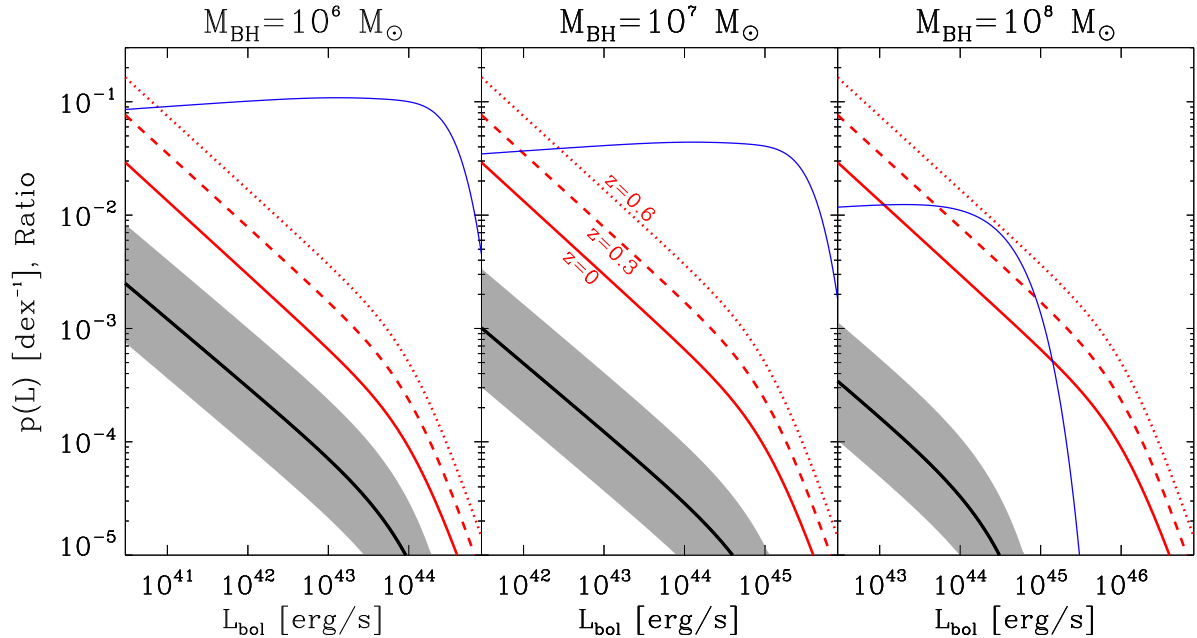}
\caption{The probability for a galaxy hosting a black hole of mass $M_{\rm BH}$
to have bolometric luminosity in the range $L_{\rm bol}$ to $L_{\rm bol}+d\log L_{\rm bol}$ are shown for three values of $M_{\rm BH}$ ($10^6 M_{\odot}$, left; $10^7 M_{\odot}$, center; $10^8 M_{\odot}$, right).
In each panel the three red lines represent the overall AGN probability function derived empirically from AGN X-ray surveys by \citet{aird12}, for three redshift values ($z=0$, solid; $z=0.3$, dashed; and $z=0.6$, dotted). the black solid line is the rate for TDE computed averaging Eq.~\ref{eq:prob} over a \citet{kroupa01} IMF from 0.3 to 30 $M_{\odot}$. The grey band represents a one-order-of-magnitude uncertainty in the TDE rates. In each plot the thin blue solid line shows the
ratio of the black to red solid lines, i.e. the $z=0$ fraction of luminous galactic nuclei powered by tidal disruptions of main-sequence stars.}
\label{fig:rates_lum}
\end{figure*}

\subsection{Tidal disruption flares vs. gas accretion in galactic nuclei}
\label{sec:rates}
We discuss here a simple unified model for the stochastic `activation' of a nuclear black hole, be it via tidal disruption flares (`TDE') 
or accretion of gas parcels from 
the galactic ISM (`ACC'). Without loss of generality, and for the sake of simplicity, we assume here that individual accretion events are all characterized by 
a light-curve composed of two main phases: a `peak' phase, where the luminosity rises rapidly and saturates at a peak value for a time $\tau_i(M_{\rm BH},z)$ 
(where, here and in the following, the index $i={\rm ACC},{\rm TDE}$, depending on the case we are interested in), 
and a `decline' phase, where the light-curve follows a power-law 
decline $L\propto t^{-\gamma_i}$. While this is now a well tested theoretical framework for TDE, as we have discussed above, 
it has also been shown by a number of authors that AGN light-curves of this sort do indeed reproduce generic features of the active galactic nuclei distributions
such as luminosity functions, accretion rate distributions, clustering properties, etc. \citep{yu04,hopkins05,yu08,bonoli09}. We 
further define the rates of nuclear activation as 
$\Gamma_i(M_{\rm BH},z)$. It is straightforward to show that, under these simplified circumstances, the probability of a galaxy hosting a black hole of mass $M_{\rm BH}$
which has bolometric luminosity between $L_{\rm bol}$ and $L_{\rm bol}+d\log L_{\rm bol}$ is given by:
\begin{equation}
\label{eq:prob}
p_i(L_{\rm bol})=\frac{1}{\gamma_i}\Gamma_i \tau_i \left(\frac{L_{\rm bol}}{L_{\rm peak}}\right)^{-1/\gamma_i}f(L_{\rm bol}/L_{\rm peak}),
\end{equation}
where $f(L_{\rm bol}/L_{\rm peak})$ is a 'cut-off' function which is $\approx$ unity for $L_{\rm bol} \ll L_{\rm peak}$, and rapidly declines for 
$L_{\rm bol}>L_{\rm peak}$ (typically exponentially).

In the case of tidal disruption of stars ($i={\rm TDE}$), we can write an approximate solution neglecting any redshift dependence, 
fixing $\gamma_{\rm TDE}=5/3$, 
$\tau_{\rm TDE}=t_{\rm peak}(M_{\rm BH},M_{*},\beta)$ (Eq.~\ref{eq:t_peak}), and 
$L_{\rm peak}=\epsilon \dot M_{\rm peak}(M_{\rm BH},M_{*},\beta) c^2$ (Eq.~\ref{eq:mdot_peak}). For simplicity, we discuss here only 
tidal disruptions of main-sequence stars, and refer the reader to the more accurate, and comprehensive work of \citet{Macleod13}, who have computed the
contribution to $p_{\rm TDE}(L_{\rm bol})$ of giant stars, either disrupted, or undergoing repeated, episodic, mass-transfer events (`spoon feeding'), 
which tend to dominate at very low luminosities.
As we discussed in the introduction, there is substantial uncertainty in the true rate of tidal stellar disruption in galactic nuclei, with current observationally-determined
rates lying factors of a few to ten below the theoretical expectations. Here, for illustration purposes, we adopt a generic (i.e. independent of black hole mass) rate
of $\Gamma_{\rm TDE}=3\times 10^{-5} M_{*}^{-1/3} R_{*}^{1/4}$, but consider, in the following, a one-order-of-magnitude allowed range around such a value. The resulting probability distributions, $p_{\rm TDE}(L_{\rm bol})$, are shown with a solid black line in Fig.~\ref{fig:rates_lum}, 
for three different values of $M_{\rm BH}$.

In the case of gaseous accretion, most attempts to reproduce the AGN luminosity function  have implied peak luminosities of the order of the Eddington luminosity. However, both the triggering rate $\Gamma_{\rm ACC}(M_{\rm BH},z)$ and the peak duration $\tau_{\rm ACC}(M_{\rm BH},z)$ are essentially unknown.
In fact, it is unlikely that a single mechanism could provide a complete description of (gaseous) AGN triggering. 
Nevertheless, multi-wavelength surveys of AGN have been used, in recent years, to provide empirical constraints on $p_{\rm ACC}(L_{\rm bol})$, by studying in a uniform way the 
properties of the host galaxies of complete AGN samples typically selected in the X-ray band, to minimize the `obscuration' bias \citep{bongiorno12,aird12}. 
Interestingly enough, \citet{aird12} showed that the $p_{\rm ACC}(L_{\rm bol})$ is independent of black hole mass, but strongly evolving with redshift 
$p(L_{\rm bol})\propto (1+z)^{3.5}$, a result confirmed by \cite{bongiorno12}, both qualitatively and quantitatively.
We make use here of the analytic solution of \citet{aird12}, who expressed $p_{\rm ACC}(L_{\rm bol})$ as a broken power-law, 
with slope $1/\gamma_{\rm ACC}=0.65$ and a
rapid decline above the Eddington luminosity. 

The very fact that, on average, the overall AGN population is consistent with $\gamma_{\rm ACC}\simeq 1.5$ is interesting,
as this is not only a steeper decline than predicted by theoretical models of viscously evolving accretion discs \citep[$\gamma_{\rm ACC} \approx 1.1 - 1.2$, ][]{cannizzo90,yu04}, but also shallower than the `self-regulated' light-curve of the quasar-feedback dominated major mergers simulated by \citet{hopkins05}, which
predict $\gamma_{\rm ACC} \approx 2$. Instead, the empirical results of \citet{aird12} indicate that, in an average sense, the light-curves of AGN do resemble 
those of tidal disruption flares (we have $\gamma_{\rm TDE}\simeq 1.66$ and $\gamma_{\rm ACC}\simeq 1.5$); the difference in the overall probability functions between the two populations must then be the outcome of widely different 
triggering rates, event durations and (to a lesser extent) peak luminosities.

Fig.~\ref{fig:rates_lum} shows, for three different values of $M_{\rm BH}$ the probability of hosting an active nucleus with luminosity 
$L_{\rm bol}$ 
(per unit logarithmic luminosity interval) for the full AGN population, as parametrized by \citet{aird12} (red lines), and for main-sequence stars TDE 
(solid black line with grey band uncertainty), where we have averaged over a \citet{kroupa01} IMF, and taken $\beta={\rm min}[1,\beta_{\rm max}(M_{\rm BH},M_{*})]$, where 
$\beta_{\rm max}(M_{\rm BH},M_{*})$ is the maximum value of the penetration factor still compatible with a tidal disruption 
(i.e. that value for which $R_{\rm T}/R_{\rm S}=1$, see Eq.~\ref{eq:rt_rs} above). The plot also shows, in each panel, the ratio of the TDE probability function to the
$z=0$ AGN probability function, which represent the fraction of active nuclei (of a given mass) at any time which are in fact powered by the disruption 
of a main sequence star. 

\begin{figure}
\includegraphics[width=0.5\textwidth,clip]{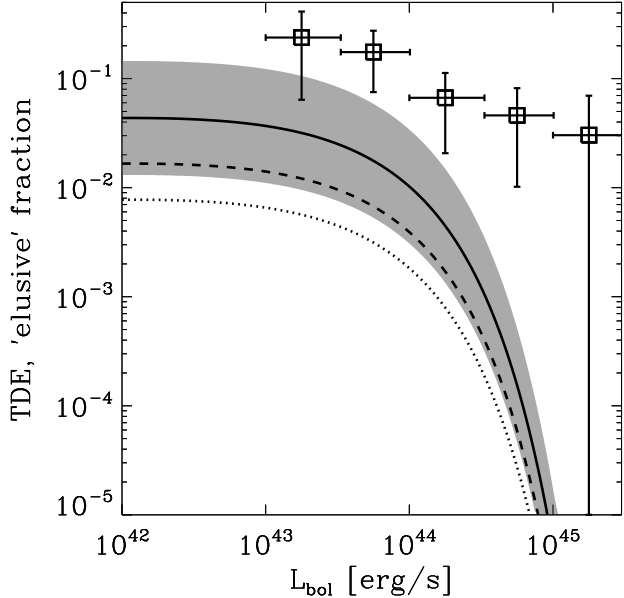}
\caption{The overall fraction of active galactic nuclei powered by TDE at $z=0$ is shown as function of nuclear bolometric luminosity (black solid line), with
a grey band representing a one-order-of-magnitude uncertainty in the TDE rates. The dashed and dotted lines show the evolution of the TDE fraction at $z=0.3$ 
and $z=0.6$, respectively, if the TDE rates are assumed to be unchanged, while the total AGN triggering rate evolves as observed empirically by \citet{aird12} 
($p_{\rm ACC}\propto (1+z)^{3.5}$). The empty squares are the observed fractions of ``elusive'' AGN  among 
all X-ray selected AGN in a highly spectroscopically complete sample (with flux $F_{\rm 2-10 keV}> 2\times 10^{-14}$ \cgs)
in the redshift range $0.1<z<0.6$ from the BOSS follow-up of the XMM-XXL survey (Menzel et al., in prep.).}
\label{fig:tde_frac}
\end{figure}

We can also weigh the probability functions $p_i(L)$ with the local black hole mass function (BHMF) and compute the total fraction of TDE-powered nuclei as a function of nuclear bolometric luminosity for a volume limited sample of local AGN. We adopt the BHMF of \citet{merloni08}, and show the results in Fig.~\ref{fig:tde_frac}. A similar calculation was performed by \citet{milos06}, who compared
TDE volumetric rates with X-ray AGN luminosity functions, and reached similar conclusions as those we can draw here: the fraction of TDE in current AGN samples
is highly uncertain, due to the still poorly known intrinsic rates, but, in the range $10^{42}<L_{\rm bol}<10^{44}$\,erg\,s$^{-1}$ could be as high as 10\%. 
The combined effect of the Eddington limit and of the fact that bigger black holes would swallow most main-sequence stars whole, produces a rapid drop of the
TDE fraction at bolometric luminosities above $10^{45}$\,erg\,s$^{-1}$, i.e. into the quasar regime. The exact behaviour of the TDE fraction at such high luminosities, however, critically depends on the SMBH
spin distribution, as rapidly spinning black holes can tidally disrupt stars even if they are very massive (see the right panel of Fig.~\ref{fig:good} above). On the other hand, a proper general-relativistic treatment
of TDE could modify the $p(L_{\rm bol})$ even at low $M_{\rm BH}$, where the Lense-Thirring precession of the stellar debris may cause severe delays in the onset of accretion, 
increasing $t_{\rm peak}$ and reducing $L_{\rm peak}$ \citep{guillochon15}.

As a term of reference, we also show in Fig.~\ref{fig:tde_frac} the observed fraction of all X-ray selected AGN in a highly spectroscopically complete sample in the redshift range $0.1<z<0.6$ (from the BOSS follow-up of the XMM-XXL survey, Menzel et al., in prep.), which are optically classified as 'elusive', or 'X-ray Bright and Optically Normal Galaxies' (XBONGS), i.e. lacking any spectroscopic signature of AGN emission \citep[see e.g.][]{comastri02,smith14,pons14}. At these luminosities,
about 6-20\% of all X-ray selected AGN are 'elusive', and we suggest that a non-negligible fraction of them are powered by 
TDE which have faded away by the time the
optically spectroscopic follow-up was performed (3 to 10 years after the X-ray data were taken in this field). Further clues on the nature of elusive AGN, and on the true fraction of TDEs among them could come from a detailed study of the IR emission from those sources, which could reveal either the 
presence of a deeply obscured AGN or the transient echo of the TDE emission. Unfortunately, no systematic, deep NIR/MIR data are available for the full X-ray selected XMM-XXL sample.

\section{Conclusions}
\label{sec:conclusions}
We have analysed in detail the light-curve and spectra of the flare observed between 1998 and 2005
in SDSS~J0159+0033, a massive star-forming galaxy in the SDSS Stripe 82 field. The most striking features of the system are: (i) a rapid increase in the X-ray and optical/UV luminosity of the source by at least one order of magnitude, and (ii) the dramatic change of its optical spectrum
over a ten year timescale. The object changed from a typical type-1 AGN emission, with prominent broad emission lines of H$\alpha$ and H$\beta$ and a blue continuum over-imposed on a galaxy spectrum, to a galaxy-dominated continuum. This latter spectrum shows strong narrow emission lines, which could have been ionized by either newly formed stars or (more likely) by past AGN activity, and, in addition, a weak broad H$\alpha$ emission line, suggesting low-level of ongoing activity a few years after the peak. 

We have studied the light-curve of the object over the last $\sim$30 years, in optical/UV and X-rays, including a well-sampled period coincident
with the flare decay from the SDSS Stripe 82 survey. The variable nuclear emission, both in X-rays and in the optical bands, is well
described by a single flaring episode, with fast rise and a power-law decay with exponent $n_{\infty}=-1.59\pm0.05$, consistent with detailed
predictions 
of hydrodynamical models of tidal disruption flares \citep{guillochon14}. 
The overall fluence of the event indeed suggests that the amount of accreted mass is  a substantial fraction (at least about 20\%) of a massive star. 
The decay light-curve also shows clear structure, with a re-brightening flare observed about four rest-frame years after the peak, which
has a fluence of about 14\% of the total fluence of the event.

We have used late-time photometric and spectroscopic observations of the source to build a host galaxy model, that we used to carefully analyse the nuclear (host-subtracted) light-curve of the flare in all five SDSS bands. This has allowed us to study the SED evolution of the flare, which we found, consistent with most
optically-selected TDEs observed to date, to display very little temperature (colour) evolution, as predicted by models in which the accretion energy released in the event is reprocessed by dense, large-scaleheight material \citep{guillochon14}.
Assuming standard (and conservative) bolometric corrections, this is among the most luminous non-beamed tidal disruption flares discovered so far, 
and the only one produced by a black hole as massive as $10^8$ solar masses. Perhaps due to the serendipitous nature of its discovery, it is also one of the few cases in which both optical and X-ray emission contribute substantially to the bolometric luminosity of the event, suggesting that accretion onto black holes from
tidally disrupted stellar debris can also give rise to powerful Comptonising coronae.
In most previous cases,
optically selected TDE had low photospheric temperatures and were X-ray faint, while X-ray selected TDE tended to be hot and optically faint
\citep[see e.g. Fig. 4 in][]{gezari12b}, clearly suggesting we are barely scratching the surface in our understanding of TDE selection effects.

If the continuum evolution and the overall energetics of the event strongly suggest a TDE interpretation, the emission line spectra tell a more complicated story.
The broad line region, which responded almost instantaneously to the strong illumination from the central flaring source in 2000, is too extended 
(about 30 light days) and massive (more than hundred solar masses) to have been 
created by the debris of a disrupted star, and must have been left, in the dark, from previous episodes of more prolonged accretion onto the central
 black hole. Also, the narrow emission lines, which we see unchanged between 2000 and 2010, possibly indicate that the central black hole was active as little as $10^4$ years ago, although it is not straightforward to estimate exactly at which level, as the narrow emission lines could be also partly excited by the 
powerful star-formation observed in this galaxy.

Such a complex interplay must not be too exceptional, also given the fact that this TDE was discovered in a relatively small parent sample. 
Within a simple, common framework in which both TDE and gas accretion onto a SMBH are modelled as stochastic
events with homologous light-curves, but widely differing triggering rates and durations, we showed how the available empirical and theoretical constraints suggest
that TDE may represent a substantial fraction (between $\approx$ 1\% and 10\%) of all 
putative AGN in ``snapshot'' surveys, at least for bolometric luminosities $L_{\rm bol}<10^{45}$ erg\,s$^{-1}$.
In particular, we expect that tidal disruption flares should be responsible for at least a fraction of those X-ray selected AGN, which, upon later
follow-up analysis, do not reveal any evidence for AGN-induced signatures in their long-wavelength spectra 
\citep[such as the so-called XBONGS,][]{comastri02,smith14}. Systematic re-observations of previously classified broad line AGN has only 
been done for small samples of nearby AGN; \citet{scott14} have indeed reported a considerable number (8/97) of `changing look' sources 
\citep[see also][for the results of a recent AGN spectroscopic monitoring campaign]{barth15}.
We are, indeed, used to think of AGN as being either type 1 (un-obscured) or type 2 (obscured), based on the properties of their optical 
emission lines. However, on timescales longer than the duration, $\tau_i$, of episodes of high-luminosity (`flares'), optical emission line classification is a 
transient variable. This is particularly important for tidal disruption events, of course, which have flare durations of a few years, at most, but could be 
relevant for other forms of black hole fuelling, too.

A few lessons can be learned from this work: probably the most compelling is that an accurate observational determination of TDE rates requires searches that do
not {\it a priori} exclude objects previously classified as AGN, and needs to fully account for the exact time when such classification was firstly made. Also,
we showed that TDE confirmation is far from straightforward, and requires well sampled optical light-curves, and would be aided by repeated, early spectroscopic observations, too. 
LSST \citep{ivezic08}, in combination with flexible, rapid responding spectroscopic survey instruments \citep[such as 4MOST, for example,][]{dejong14}, 
will be instrumental for our understanding of nuclear variability and AGN triggering.
In addition, deep and wide X-ray observation with the right cadence have the advantage that the TDE-produced light becomes confused with the host galaxy stellar background at much later stages at these energies, thus allowing flares to be identified with less well-sampled light-curves and on longer timescales. Thanks to this fact, 
the upcoming eROSITA on SRG \citep{merloni12} could yield
as many as two TDE candidates alerts per day \citep{khabibullin14}, requiring substantial dedicated resources to be mobilized in order to follow most of them up.

\section*{Acknowledgments}
We thank Zhu Liu, Marie-Luise Menzel, Enrico Ramirez-Ruiz, Chris Reynolds, Paola Santini, Yue Shen and Nicholas Stone for useful discussions. 
JG, AM and MS acknowledge support from 
the DFG cluster of excellence ``Origin and structure of the universe'' ({\tt www.universe-cluster.de}). 
GP acknowledge support via an EU Marie Curie Intra-European 
fellowship under contract no. FP-PEOPLE-2012-IEF- 331095 and  the Bundesministerium
f{\"u}r	Wirtschaft und Technologie/Deutsches Zentrum f{\"u}r Luft- und 
Raumfahrt (BMWI/DLR, FKZ 50 OR 1408) and the Max Planck Society.

Part of the funding for GROND (both hardware as well as personnel) 
    was generously granted from the Leibniz-Prize to Prof. G. Hasinger 
    (DFG grant HA 1850/28-1).

This research made use of the cross-match service provided by CDS,
Strasbourg.

Funding for the SDSS and SDSS-II has been provided by the Alfred P. Sloan Foundation, the Participating Institutions, the National Science Foundation, the U.S. Department of Energy, the National Aeronautics and Space Administration, the Japanese Monbukagakusho, the Max Planck Society, and the Higher Education Funding Council for England. Funding for SDSS-III has been provided by the Alfred P. Sloan Foundation, the Participating Institutions, the National Science Foundation, and the U.S. Department of Energy Office of Science. The SDSS web site is {\tt http://www.sdss.org/}.

    The SDSS was managed by the Astrophysical Research Consortium for the Participating Institutions. The Participating Institutions are the American Museum of Natural History, Astrophysical Institute Potsdam, University of Basel, University of Cambridge, Case Western Reserve University, University of Chicago, Drexel University, Fermilab, the Institute for Advanced Study, the Japan Participation Group, Johns Hopkins University, the Joint Institute for Nuclear Astrophysics, the Kavli Institute for Particle Astrophysics and Cosmology, the Korean Scientist Group, the Chinese Academy of Sciences (LAMOST), Los Alamos National Laboratory, the Max-Planck-Institute for Astronomy (MPIA), the Max-Planck-Institute for Astrophysics (MPA), New Mexico State University, Ohio State University, University of Pittsburgh, University of Portsmouth, Princeton University, the United States Naval Observatory, and the University of Washington.

SDSS-III was managed by the Astrophysical Research Consortium for the Participating Institutions of the SDSS-III Collaboration including the University of Arizona, the Brazilian Participation Group, Brookhaven National Laboratory, Carnegie Mellon University, University of Florida, the French Participation Group, the German Participation Group, Harvard University, the Instituto de Astrofisica de Canarias, the Michigan State\/Notre Dame\/JINA Participation Group, Johns Hopkins University, Lawrence Berkeley National Laboratory, Max Planck Institute for Astrophysics, Max Planck Institute for Extraterrestrial Physics, New Mexico State University, New York University, Ohio State University, Pennsylvania State University, University of Portsmouth, Princeton University, the Spanish Participation Group, University of Tokyo, University of Utah, Vanderbilt University, University of Virginia, University of Washington, and Yale University
. 

\footnotesize{
\bibliographystyle{mn2e_mod}
\bibliography{tde}
}

\appendix

 \section{Photometric calibration of DSS Palomar and UK-Schmidt photographic data against SDSS}
 \label{appendix_dss}
In order to convert and calibrate the pre-1998 photographic imaging data-points from the DSS 
\citep[as shown in Fig.~\ref{fig:long_term_lc}; from GSC 2.3.2 catalogue,][]{lasker08} into the SDSS photometric system, 
we cross-correlated point-like detections from the GSC 2.3.2 catalogue lying within 2 degrees from SDSS J0159+0033 with SDSS
photometric objects (treated as a truth sample) in order to measure the
bandpass conversion formulae and residual scatter for the photographic
plate photometry. To relate the SDSS {\it g', r', i'} bands to the observed $j$, $F$, $N$ and $V$ magnitudes, 
we used fitting formulae that are linear in SDSS colour and quadratic in the
photographic magnitude, but the quadratic term turns out to be negligible in all but the $F$-band conversion. 
We obtained the following expressions for the SDSS magnitudes:

\begin{equation}
 g'(j,g'-r') = 0.38811 + 0.96837 j -0.089848 (g'-r'-1)
\end{equation}

\begin{equation}
r'(F,g'-r') = -8.8646 + 1.4706 F + 0.10263 (F-20)^2 + 0.10345 (g'-r'-1)
\end{equation}

\begin{equation}
i'(N,r'-i') = 1.2587 + 0.94722 N + 0.2751 (r'-i'-0.5)
\end{equation}

\begin{equation}
r'(V,g'-r') = 0.27671 + 0.96792 V - 0.22833 (g'-r'-1)
\end{equation}

We adopted as a measure of uncertainty the standard deviation of the residuals of
points lying within $\pm 0.5$  photographic magnitudes of
SDSS J0159+0033 (and with a difference between observed and predicted SDSS magnitude $<3$ mag).
Finally, we assume that $g-r=1.0$ and $r-i = 0.5$ (AB mag) when converting the
photographic photometry of J0159+0033 (which means that the colour terms disappear for SDSS J0159+0033).
Table~\ref{tab:photo} lists the derived photometric points.

\begin{table*}
\caption{Conversion of DSS photographic plates magnitudes to SDSS photometric system for optical imaging  data of SDSS~J0159+0033 taken before 1998. 
The source for the observation dates for the photographic plates was taken from {\tt https://archive.stsci.edu/cgi-bin/dss\_plate\_finder}.}
\label{tab:photo}
\begin{tabular}{lcccc}
\hline
Epoch & GSC Band & Mag  & SDSS band  & Mag (AB)  \\
\hline
1983-09-09 & $V$ & $19.52 \pm 0.79$ & $r'$ & $19.17 \pm 0.36$ \\
1991-11-02 & $N$ & $18.66 \pm 0.47$ & $i'$ & $18.93 \pm 0.27$ \\
1992-10-20 & $j$ & $19.94 \pm 0.36$ & $g'$ & $19.70 \pm 0.22$ \\
1996-10-14 & $F$ & $18.78 \pm 0.40$ & $r'$ & $18.91 \pm 0.26$ \\
\hline
\end{tabular}
\end{table*}

\bsp
\label{lastpage}

\end{document}